\documentclass[10pt,journal,compsoc]{IEEEtran}
\pdfoutput=1 
\usepackage[switch]{lineno}

\usepackage{tabu}                      
\usepackage{booktabs}                  
\usepackage{amsmath}

\usepackage{array}

\newcolumntype{L}[1]{>{\raggedright\let\newline\\\arraybackslash\hspace{0pt}}m{#1}}
\newcolumntype{C}[1]{>{\centering\let\newline\\\arraybackslash\hspace{0pt}}m{#1}}
\newcolumntype{R}[1]{>{\raggedleft\let\newline\\\arraybackslash\hspace{0pt}}m{#1}}

\ifCLASSOPTIONcompsoc
  \usepackage[nocompress]{cite}
\else
  \usepackage{cite}
\fi
\ifCLASSINFOpdf
\else
\fi
\usepackage[normalem]{ulem}
\usepackage[usenames,dvipsnames,svgnames,table]{xcolor}
\usepackage{ifthen}

\newboolean{final}
\setboolean{final}{true}

\newenvironment{SUBENVcomment}[3]{\color{#1}(#2:#3)}{\color{black}}
\newenvironment{SUBENVhighlight}[1]{\color{#1}}{\color{black}}

\definecolor{author1}{rgb} {0.0,0.6,0.8}
\definecolor{author2}{rgb} {0.9,0.25,0.0}
\definecolor{author3}{rgb} {0.0,0.5,0.0}
\definecolor{author4}{rgb} {0.6,0.0,0.8} 
\definecolor{author5}{rgb} {0.9,0.25,0.47}
\definecolor{cfixme}{rgb}  {0.9,0.2,0.2} 
\definecolor{cresume}{rgb}  {0.9,0.2,0.9} 
\definecolor{cadded}{rgb}  {0.0,0.9,0.0} 

\newcommand{\comment}[1]{}

\newcommand{\commentary}[4]{\ifthenelse{\boolean{final}}{}{\begin{SUBENVcomment}{#1}{#2}{#3}~#4\end{SUBENVcomment}}}

\newcommand{\fixme}[1] {\ifthenelse{\boolean{final}}{~#1}{\begin{SUBENVhighlight}{cfixme}~#1\end{SUBENVhighlight}}}
\newcommand{\resume}[1] {\ifthenelse{\boolean{final}}{~#1}{\begin{SUBENVhighlight}{cresume}~#1\end{SUBENVhighlight}}}
\newcommand{\added}[1] {\ifthenelse{\boolean{final}}{~#1}{\begin{SUBENVhighlight}{cadded}~#1\end{SUBENVhighlight}}}
\newcommand{\instead}[1]

\newcommand{\delete}[1]{\ifthenelse{\boolean{final}}{   }{\sout{#1}}}

\newcommand{\replace}[2]{\ifthenelse{\boolean{final}}{#1}{\added{#1}\delete{#2}}}

\newcommand{\anovaETAbody}[6]{{$F(#1,#2)$\,$=$\,$#3$, $p$\,$#4$\,$#5$, $\eta_{p}^{2}$\,$=$\,$#6$}}

\usepackage{paralist}
\newenvironment{itemize*}%
  {\begin{compactitem}%
    \setlength{\plitemsep}{2pt}}%
  {\end{compactitem}}

 \newenvironment{enumerate*}[1][]%
  {\begin{compactenum}[#1]%
    \setlength{\plitemsep}{1pt}}%
  {\end{compactenum}}

\usepackage{soul}	
\usepackage[normalem]{ulem}

\newboolean{doHighlight}
\setboolean{doHighlight}{true}
\ifthenelse{\boolean{doHighlight}}
{
\soulregister\cite{7}
\soulregister\shortcite{7}

\usepackage[colorinlistoftodos]{todonotes}
\usepackage{setspace}
\setlength{\marginparwidth}{1.5cm}
}
{

\usepackage[colorinlistoftodos, disable]{todonotes}
}

\usepackage{eurosym}

\begin{document}
%

\title{Virtual Co-Embodiment: \\ Evaluation of the Sense of Agency \\ while Sharing the Control of a Virtual Body}

\title{Virtual Co-Embodiment: Evaluation of the Sense of Agency while Sharing the Control of a \\ Virtual Body among Two Individuals}


%
%
%
%

\author{Rebecca Fribourg*, 
        Nami Ogawa*, 
        Ludovic Hoyet, Ferran Argelaguet,\\ Takuji Narumi, Michitaka Hirose and~Anatole L\'ecuyer%
\IEEEcompsocitemizethanks{
\IEEEcompsocthanksitem *: Both authors have contributed equally to this work and are listed in an alphabetical order.\protect\\
    \IEEEcompsocthanksitem Rebecca Fribourg, Ludovic Hoyet, Ferran Argelaguet and Anatole L\'ecuyer: Inria, Univ Rennes, CNRS, IRISA, France.\protect\\
  \IEEEcompsocthanksitem Takuji Narumi: The University of Tokyo / JST PREST\protect\\
    \IEEEcompsocthanksitem Nami Ogawa and Michitaka Hirose: The University of Tokyo}
\thanks{\copyright~2020 IEEE. Personal use of this material is permitted. Permission from IEEE must be obtained for all other uses, in any current or future media, including reprinting/republishing this material for advertising or promotional purposes, creating new collective works, for resale or redistribution to servers or lists, or reuse of any copyrighted component of this work in other works.}
}

%
%

\markboth{Journal of \LaTeX\ Class Files,~Vol.~XX, No.~X, XXX~XXXX}%
{Shell \MakeLowercase{\textit{Fribourg et al.}}: Virtual Co-Embodiment: Evaluation of the Sense of Agency while Sharing the Control of a Virtual Body among Two Individuals}

\makeatletter
 \let\old@ps@headings\ps@headings
 \let\old@ps@IEEEtitlepagestyle\ps@IEEEtitlepagestyle
 \def\confheader#1{%
 \def\ps@IEEEtitlepagestyle{%
 \old@ps@IEEEtitlepagestyle%
 \def\@oddhead{\strut\hfill#1\hfill\strut}%
 \def\@evenhead{\strut\hfill#1\hfill\strut}%
 }%
 \ps@headings%
 }
 \makeatother

\confheader{%
\parbox{18.5cm}{\center{\scriptsize{This is the author's version of an article that has been published in this journal. Changes were made to this version by the publisher prior to publication.
\\The final version of record is available at http://dx.doi.org/10.1109/TVCG.2020.2999197}}}
 }


\IEEEtitleabstractindextext{%

\begin{abstract}
In this paper, we introduce a concept called ``virtual co-embodiment'', \added{which enables a user to share their virtual avatar with another entity (e.g., another user, robot, or autonomous agent)}. We describe a proof-of-concept in which two users can be immersed from a first-person perspective in a virtual environment and can have complementary levels of control (total, partial, or none) over a shared avatar. In addition, we conducted an experiment to investigate the influence of users' level of control over the shared avatar and prior knowledge of their actions on the users' sense of agency and motor actions. The results showed that participants are good at estimating their real level of control but significantly overestimate their sense of agency when they can anticipate the motion of the avatar. Moreover, participants performed similar body motions regardless of their real control over the avatar. The results also revealed that the internal dimension of the locus of control, which is a personality trait, is negatively correlated with the user's perceived level of control. The combined results unfold a new range of applications in the fields of virtual-reality-based training and collaborative teleoperation, where users would be able to share their virtual body.
\end{abstract}

\begin{IEEEkeywords}
Virtual Embodiment, Sense of Agency, Avatars, Virtual Reality, User Experimentation
\end{IEEEkeywords}}

\maketitle

\IEEEdisplaynontitleabstractindextext

%
\IEEEpeerreviewmaketitle


%
%
%
%

\begin{figure*}
\centering
  \includegraphics[width=0.58\linewidth]{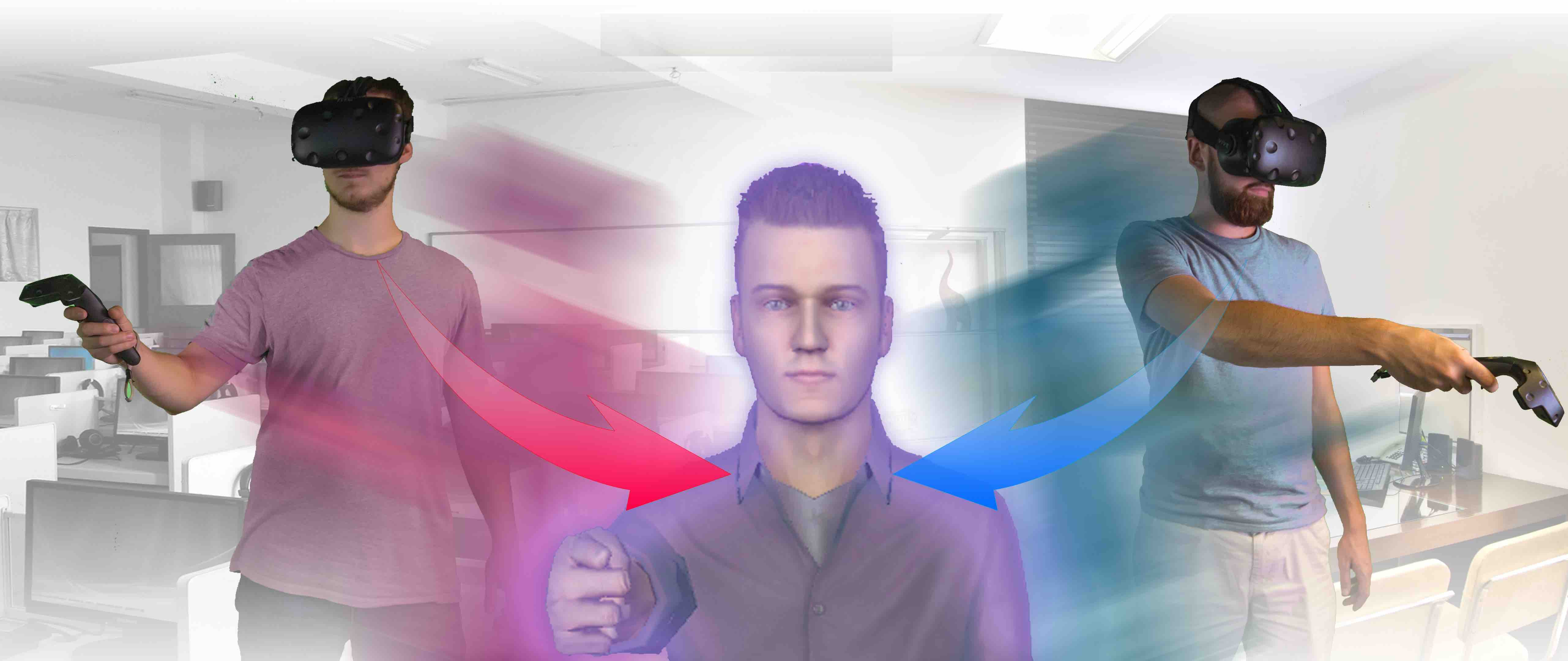}
  \hfill{}
  \includegraphics[width=0.36\linewidth]{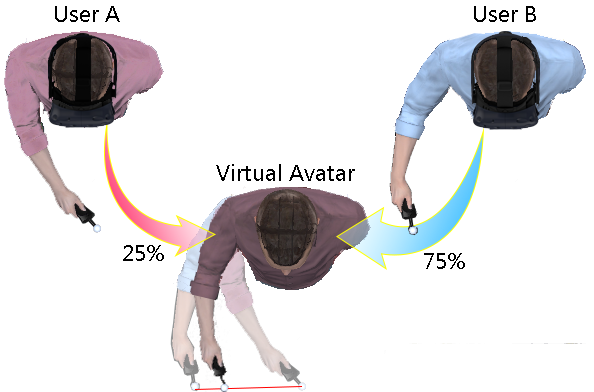}
  \caption{Our ``Virtual Co-Embodiment'' experience enables a pair of users to be embodied simultaneously in the same virtual avatar (Left). The positions and orientations of the two users are applied to the virtual body of the avatar based on a weighted average, e.g., ``User A'' with 25\% control and ``User B'' with 75\% control over the virtual body (Right).}
  \label{fig:teaser}
\end{figure*}

\IEEEraisesectionheading{\section{Introduction}\label{sec:intro}}

The emerging use of self-avatars in virtual reality (VR) has uncovered numerous novel possibilities to explore the relation between body and mind \cite{Kilteni2012,Hoyet2016}. Indeed, avatars in VR enable original experiences as they can be altered and controlled in numerous ways. For example, it is possible to be embodied in avatars with a different gender \cite{8260949} or with morphological changes such as possessing a hand with six fingers \cite{Hoyet2016}. Such experiences have helped 
to better understand how users perceive their virtual representation in VR and to explore how users are willing to accept a virtual body that differs from their own in terms of visual aspect and control schemes. 
%
%
Furthermore, the need to collaborate in VR reinvigorates research interests in developing new ways for users to collaboratively interact in a virtual environment (VE), especially through their avatars~\cite{Latoschik2017, 10.1371/journal.pone.0189078}. More specifically, several VR shared experiences, in which two individuals can share a first-person view, have been explored, especially in terms of gesture training or collaborative teleoperation~\cite{Yang2002, Kawasaki2010}. However, previous studies on this topic did not involve scenarios in which several users could be embodied in
the same avatar.


\added{In this paper, we introduce a new concept, termed ``virtual co-embodiment.'' While the concept of ``co-embodiment'' has been recently defined outside of the scope of VR~\cite{Luria:2019:RCE:3322276.3322340}, this is the first study to the best of the authors' knowledge to define ``virtual co-embodiment'' as a situation that enables a user and another entity (e.g., another user, robot, or autonomous agent) to be embodied in the same virtual avatar. }
%
%
Such a situation raises the question about how sharing a virtual body influences ones' perception and actions in the VE. Potential applications of this new concept range from VR-based motion training to collaborative teleoperation, e.g., to efficiently transfer physical skills from an expert to a novice, or to enable the simultaneous control of a robot by two experts as if the robot was their actual body.
In such scenarios, it is therefore important to maintain the feeling of control for both users so that they have the impression that they are controlling the avatar in the same manner that they would control their own bodies.

As a first step, this study focused on two users sharing the same virtual body. In VR, the sense of embodiment (SoE) is a theoretical framework widely used to evaluate how users perceive and accept their avatar to be their own representation in the virtual world~\cite{Kilteni2015, Longo2008}. This framework is often divided into three dimensions~\cite{Kilteni2012}: the sense of agency (SoA), sense of self-location (SoSL), and sense of body ownership (SoBO). However, owing to the particularity of the virtual co-embodiment experience, in which users share control over their virtual body, and the potential implications that sharing this control would increase the interaction capabilities of users in a VE, we decided to focus our efforts on the assessment of the SoA.
The ability to modulate the sharing of avatar control enables the possibility to assess the SoA when two users collaborate to achieve a task while embodied in the same virtual avatar. 
%
%
Previous research explored the influence of perceptual and motor mismatches over the SoA. Such studies showed that it is possible for users to feel agency toward actions they did not perform~\cite{Wegner2004}, 
and highlighted interesting insights regarding the SoA with its possible modulations, inspiring the following question: To which level can users experience a SoA over a shared virtual avatar?
%

To answer this question, we conducted a VR experiment in which 12 pairs of individuals participated. Each pair was embodied in the same shared avatar from a first-person perspective (1PP) and was asked to perform different tasks in the VE while sharing the avatar control. The control was shared by averaging the position and orientation of the hands of both users according to a predefined level of control for each user (from no-control to full-control) and by animating the avatar accordingly. 
\added{Our two main hypotheses were as follows: (1) the SoA would be positively correlated with the degree of control over the shared avatar; and (2) the SoA would be positively influenced by how much the task is potentially restricting the participant's choices.}
Overall, our results support our main hypotheses, showing that the SoA over a shared avatar is significantly dependent on both the users' level of control and freedom of movements inferred by the type of task.
Whether users possessed the same prior knowledge of the action to perform also influences the SoA. Interestingly, we observed that users tend to feel some control over the avatar even when they actually have little or no control over the virtual body, in cases where their movements are more constrained. This suggests that even with little or no control over a shared avatar, users are capable of feeling some agency toward the virtual body. Our results also reveal that the internal dimension of the locus of control (LoC), which is a user personality trait, is negatively correlated with the manner in which users feel in control. Our combined results are promising for possible applications in the fields of VR-based motion training or collaborative teleoperation, where sharing the avatar control with another user could emphasize the efficiency of previously developed systems~\cite{Yang2002, gomezhal-01228890}.

\added{The main contributions of this paper are twofold: the concept of ``virtual co-embodiment'' is introduced, and this is the first study to provide a baseline for a more in-depth analysis of virtual co-embodiment, and avatar control more generally, on human behavior and self-perception.}

\section{Related Work}
\label{sec:rw}


According to Kilteni et al.~\cite{Kilteni2012}, the SoE consists of 
SoA, SoBO, and SoSL. 
As stated earlier, SoA refers to the feeling of control (FoC) over actions and their consequences\comment{ \cite{Moore2016}}, whereas SoBO refers to one's self-attribution of body\comment{ \cite{Tsakiris2006}} and SoSL refers to one's spatial experience of being inside a body\comment{ \cite{Kilteni2012}}. In VR, the SoA can easily be elicited when the user motion is mapped onto the virtual body in real-time or near real-time~\cite{Kilteni2012}. Such visuomotor congruence can also induce SoBO~\cite{VHI}, as long as the virtual body is structurally and morphologically similar with one’s biological body~\cite{Kilteni2012}. In contrast, SoSL can be achieved by an immersion from 1PP as it is highly determined from the egocentric visuospatial perspective.
While observing a virtual-realistic body from the 1PP with congruent visuomotor cues is considered to be sufficient for inducing the SoE in VR, several studies have also demonstrated that incongruent visuomotor feedback can affect the SoE. In particular, both SoA and SoBO have been found to be reduced when a discrepancy exists between vision and motor information~\cite{VHI,Farrer2008,Franck2001a}; however, they can still be induced to some extent. For instance, Maselli and Slater~\cite{Maselli2013} showed that visual realism of the avatar favors the SoE, despite the presence of incongruent visuomotor and visuotactile cues. More recently, Kokkinara et al.~\cite{Kokkinara2016} showed that both SoBO and SoA can be induced over a virtual-body walking from a 1PP, even though participants are actually seated and only allowed head movements. Such findings suggest that participants can feel SoA and SoBO in some situations despite visuomotor discrepancies. However, how the sharing of the control of a virtual body with another user influences the SoE is unknown. Therefore, we focus on sharing the control of a virtual body in this study, and the following sections will cover related work on the SoA and shared body experiences.

\subsection{SoA}

\subsubsection{Theory}
%
%
As stated earlier, the SoA is considered as one of the components of the SoE in the VR field. However, in the fields of philosophy and psychology, the SoA is considered to form a fundamental aspect of self-awareness together with SoBO~\cite{GALLAGHER200014}. Therefore, numerous studies on SoA have been conducted in the fields of philosophy and psychology to examine human consciousness. Although the mechanisms of human consciousness are still not fully understood, two influential theoretical views have been put forward: a \textit{comparator model}~\cite{Frith2000} and \textit{retrospective inference view}~\cite{Wegner1999}.
The \textit{comparator model} suggests that the comparison between predicted and actual consequences of an action through sensorimotor processes determines the SoA~\cite{Frith2000, Blakemore1999}.
Thus, the mismatch caused by spatial and temporal distortion of movements or outcomes can attenuate the SoA~\cite{Haggard2012}.
Indeed, numerous studies have shown evidence that discrepancies between the actual movement and the corresponding visual feedback~\cite{Franck2001a,Farrer2008} or sensory outcome~\cite{Blakemore1999, Sato2005} of the action negatively affect the SoA. In comparison, \textit{retrospective inference view} emphasizes external situational cues~\cite{Wegner1999}.
According to Wegner's theory of apparent mental causation~\cite{Wegner1999}, the SoA arises if (1) an intention precedes an observed action (priority), (2) the intention is compatible with this action (consistency), and (3) the intention is the most likely cause of this action (exclusivity). Therefore, priming is often used to modulate the SoA by manipulating prior conscious thought about an outcome~\cite{Moore2009,Wenke2010, Linser2007}.
However, the SoA is increasingly recognized as being based on a combination of internal motor signals and external evidence about the source of actions and effects~\cite{Moore2009, Wegner2004, Wegner2004a}.
Thus, although spatial and temporal contiguity between one's own and observed movements are the main cues for SoA~\cite{Haggard2012,Farrer2008, Franck2001a}, higher-level cognitive processes, such as background beliefs and contextual knowledge relating to the action, also influence the induction of SoA~\cite{Moore2016, Desantis2011}.



\subsubsection{Measures}

The measurements of SoA are generally categorized as implicit and explicit measures~\cite{Moore2016}.
Implicit measures, such as sensory attenuation~\cite{Blakemore1999}, intentional binding~\cite{Haggard2002, Moore2012}, and neurophysiological markers~\cite{Jeunet2018} assess a correlation of voluntary actions about the agentic experience~\cite{Moore2016}.
Alternatively, explicit measures are based on the subjective judgments of the FoC, the authorship or attribution of the actions or their corresponding outcomes~\cite{Sato2005, Linser2007, Wenke2010, Wegner2004, Daprati1997}. Most studies have used explicit measures, especially in VR \cite{Kokkinara2015, Ma2015, Tieri2015}.
These measures are typically assessed in paradigms using button presses which produce sensory feedback~\cite{Sato2005} or simple movements~\cite{Farrer2008, Franck2001a,Wegner1999, Metcalfe2007, Nielsen1963, Daprati1997,Kokkinara2015}.
For simple movements, a moving cursor associated with a joystick~\cite{Farrer2008, Franck2001a} or mouse~\cite{Wegner1999, Metcalfe2007} and a visual feedback of hand movements through a mirror~\cite{Nielsen1963}, a TV-screen~\cite{Daprati1997}, or VR~\cite{Kokkinara2015} are often used.

Besides these experimental measurements, some personality traits have been shown to be related to SoA.
Neurological studies have shown that patients with schizophrenia tend to feel abnormal SoA, i.e., they have less ability to distinguish sensations due to self-caused actions from those due to external sources~\cite{GALLAGHER200014, Franck2001a, DeVignemont2004}. 
Indeed, schizotypy personality traits, an indicator of a predisposition to schizophrenia, have been shown to be correlated with an abnormal SoA~\cite{Asai2008}.
In addition, some studies have revealed that SoA is correlated with one's personality of LoC, which has been often used in the fields of education, health, and clinical psychology.
LoC refers to the degree to which people believe that they have control over the outcome of events in their lives, as opposed to external forces beyond their control~\cite{Levenson1981}. 
The Internal-Personal-Chance (IPC) test~\cite{LevensonIPC} is one of the measurements for LoC, indicating a person's relative standing on each of the three dimensions of internal, powerful others, and chance.
Among them, the individuals with a strong internal LoC believe events in their life are derived primarily from their own actions.
In a study including manipulations of the SoA in VR, based on the principles of priority, exclusivity, and consistency, Jeunet et al.~\cite{Jeunet2018} suggested that the internal dimension of LoC is positively correlated with participants' level of agency. 

\subsubsection{Illusory SoA}{}

As described earlier, spatial displacement or temporal delay between action and outcome attenuates the SoA~\cite{Haggard2012,Farrer2008, Franck2001a}. However, we feel illusory SoA over distorted movements as long as the displacement or delay is under the threshold. For example, a recent study using VR showed that spatial manipulations of $22\deg$ of angular offset from 1PP did not attenuate SoA~\cite{Kokkinara2015}; this showed much lower detection thresholds than previous studies without VR~\cite{Farrer2008,Franck2001a}.
In addition, illusory SoA can occur over the actions or outcomes made by someone else when there is a close match between prior intentions and subsequent actions. In a classic study by Nielsen~\cite{Nielsen1963}, participants were instructed to draw a straight line to the goal point. After some repetitions, the experimenter secretly inserted a mirror so that the participants were looking at another person's hand in a mirror. They experienced the illusory SoA and attributed the hand to their own. In Wegner and Wheatley's ``I-spy'' experiment~\cite{Wegner1999}, participants and an experimenter jointly controlled a cursor. Auditory priming of action-relevant thoughts induced illusory SoA even through the cursor was being controlled by someone else. This suggests that post-hoc judgments of SoA can easily be distorted in a joint action when the action source is ambiguous. Yokosaka et al.~\cite{Yokosaka2014} reported that when participants watched their own and another person's hand motion alternately from 1PP, they felt illusory SoA over the movement, although they were aware that they were not performing a united motion.

Moreover, illusory SoA is possible over body movements even when no actual corresponding action is being performed. In the ``helping hands`` experiment by Wegner et al.~\cite{Wegner2004}, participants watched themselves in a mirror while an experimenter standing directly behind them extended and moved his or her arms as if the participants themselves moved their arms. They reported that participants felt an illusory SoA for another person's hands when they were primed about instructions for that person's movements in advance, although they factually did not move. VR is also used to induce illusory SoA when passively observing movements of a walking avatar from 1PP~\cite{Kokkinara2016}. To summarize, in situations where individuals do not move, the action priming and movement observation from 1PP are considered to be important for illusory SoA. 
Therefore, we believe that users might experience all the three aforementioned types of illusory SoA in a virtual co-embodiment experience, as the feedback component originates partially from one's own movements and partially from someone else's movements.

\subsection{Shared Body Experiences}

Some previous studies have developed shared body experiences, e.g., two individuals sharing 1PP~\cite{Petkova2008, Kasahara2016}, kinesthetic experiences~\cite{Nishida2017}, or body representations~\cite{Petkova2008,Mazzurega2011a, Tsakiris2008, Sforza2010,Mazzurega2011a, Tajadura-Jimenez2012a}.
In particular, Petkova et al.~\cite{Petkova2008} introduced the perceptual illusion of body swapping and showed that 1PP of another person's body, in combination with the receipt of correlated multisensory information from the body, was sufficient for inducing body ownership. Mutual paralleled first-person-view-sharing systems, in which a person can observe others' first-person video perspectives as well as their own perspective in realtime, are also used in entertainment, remote collaboration, and skill transmission systems~\cite{Kasahara2016, Kawasaki2010, Poelman2012}. Other approaches have also explored the sharing of other senses, e.g., BioSync~\cite{Nishida2017} which is an interpersonal kinesthetic communication system allowing users to sense and combine muscle contraction and joint rigidity bidirectionally through electromyogram measurement and electrical muscle stimulation. Lastly, the enfacement phenomenon is a self--other merging experience, in which participants reported that morphed images of themselves and their partner contained more self than other only after synchronous multisensory stimulation on their faces~\cite{Mazzurega2011a, Tsakiris2008, Sforza2010}.

\delete{In this paper, we explore a novel concept termed ``virtual co-embodiment'', where two individuals share the same body representation and control in real-time from a 1PP, resulting in self--other bodily merging experiences.}

\section{Co-embodiment platform}
\label{sec:platform}

\begin{figure}[t]
\newcommand\imgsize{6.00cm}
\centering
\includegraphics[width=\columnwidth]{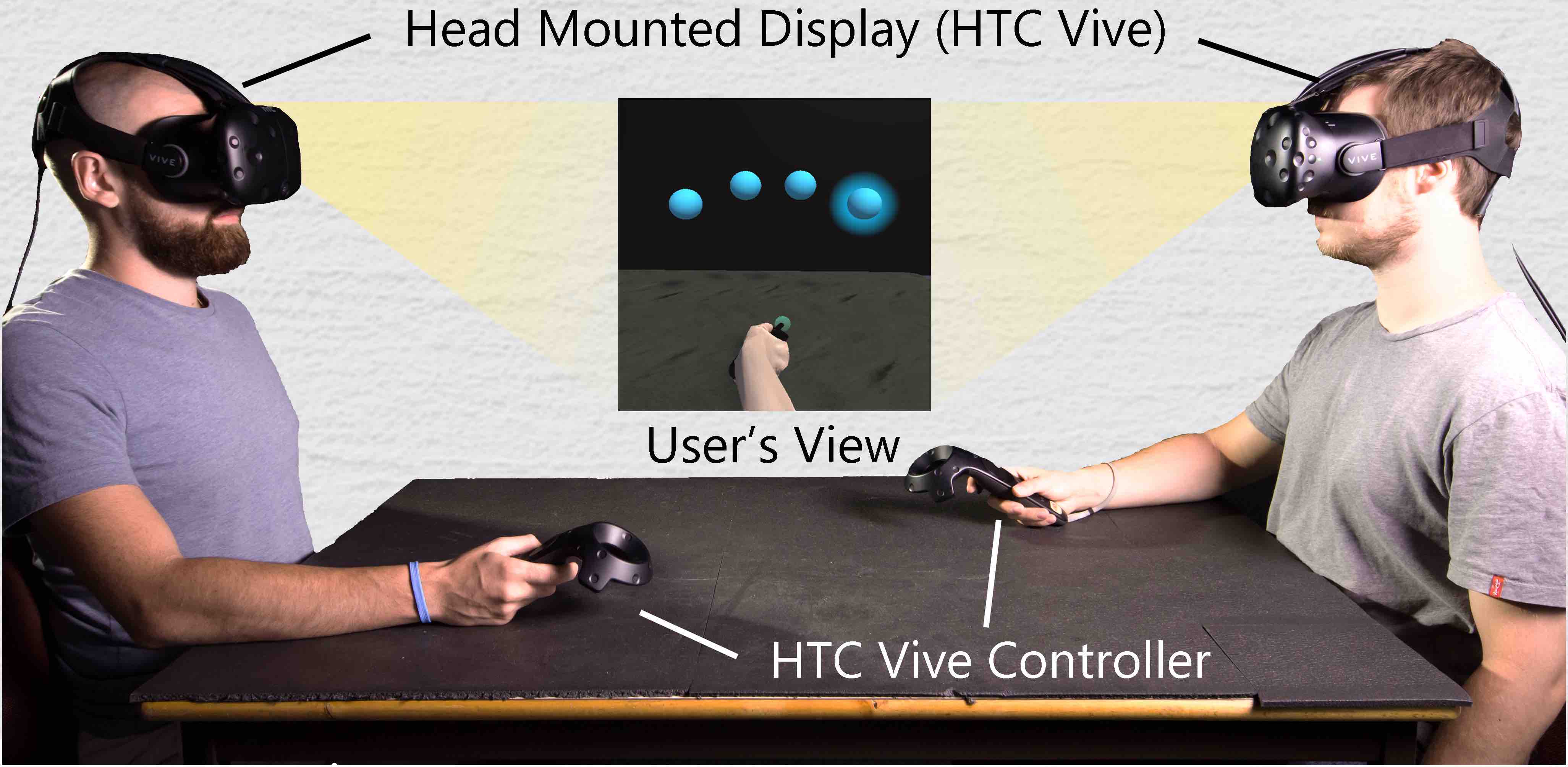}
\caption{Physical setup: two users are physically sitting in front of each other and are immersed in the same avatar from a 1PP.}
\label{fig:physicalSetUp}
\end{figure}
%


In this section, we discuss about the proposed virtual co-embodiment platform, which 
 was used to conduct an experiment, as described in Section~\ref{sec:experiment}.

The platform was developed in Unity and allows two users to share the same virtual environment and interact in real time, while being embodied in the same avatar. Our setup is based on two HTC Vive head-mounted displays (HMDs) with two HTC Vive controllers to immerse participants in the VE. The application was run on Unity 2018.1.0f2 at a constant frame rate of 90 Hz, and both computers were physically connected on the same network to minimize latency. Users were embodied in an anthropomorphic virtual avatar from a 1PP (see Figure~\ref{fig:physicalSetUp}). In the center of the tracking zone, two chairs and a table were placed, enabling users to sit in front of each other. The physical furniture had its virtual counterpart in the VE providing both a reference frame and passive haptic feedback. The VE in which users were immersed comprised an empty room. 

In terms of avatar appearance, we chose to use a realistic model in our experiment as well as immerse users in a 1PP, as these criteria were reported by recent studies to be important for enhancing the overall SoE ~\cite{Maselli2013}. As animation and control quality are known to be strongly linked to the SoA, we primarily focused on avatar animation. This was especially challenging in our case owing to the shared control of the avatar. Note that, the differentiation of avatar animation and control inputs is necessary for its computation. 

In the case of a single-user situation, the animation of the avatar depends solely on the control inputs of this user. However, in this study, the control inputs result from the combination of the inputs of two users. We therefore implemented a method that allowed the sharing of the avatar control with another user. 
As a virtual view that does not correspond to the user's own head movement could cause motion sickness, each user observed his/her own perspective in accordance with his/her head movement; the head position and orientation of the HMDs were not shared.
Regarding the controller, we computed the weighted average of the real-time position and orientation of each user's controller, and applied it to the shared avatar's controller. The weight defining the level of control could be continuously changed from 0\% to 100\%. The weighted average position and orientation were then computed by interpolating between user controller positions and orientations.

Further, we chose to focus on the animation of the arms and torso because, as stated in~\cite{Jeunet2018}, the arms and hands are the main body medium for interactions in VEs. In addition, in our setup, as users were seated on a chair, only animation for the upper body was required, which was animated through inverse kinematics using the Final IK Unity package. The Final IK computed inverse kinematics using position and rotation inputs of the head and controllers of the shared avatar, obtained through the previous shared control computation.   
Users could thus observe the same shared avatar, the movements of which, computed by inverse kinematics, would follow more or less their own hand according to their level of control at a certain time.


\section{Experiment}
\label{sec:experiment}

\subsection{Experiment Summary}


\added{We conducted an experiment, in which we explored the influence of the degree of control of an avatar shared with another person on one's own SoA. More precisely, we address the two following questions. Does the degree of shared control have an impact on one's FoC toward the avatar? Does the predictability of the avatar movement have an impact on one's FoC toward the avatar?}

In literature, the SoA was shown to largely depend on the degree of discrepancy between the predicted sensory feedback of an action and the actual outcome~\cite{Sato2005}. In addition, participants were observed to feel illusory SoA over distorted movements when the discrepancy is under a certain threshold~\cite{Kokkinara2015}. Moreover, other studies focused on situations in which participants did not move and experienced illusory SoA toward movements they did not perform, when they had prior knowledge of the action and were immersed in a 1PP~\cite{Wegner2004}. Based on these findings, we hypothesized that the level of control over the avatar shared between the two participants would influence the SoA. We also hypothesized that the freedom of movement in the task and whether both participants had the same prior knowledge of the action would also influence the SoA.

To test these hypotheses, we designed an experiment in which two participants were immersed simultaneously in a VE and were embodied in the same avatar. More precisely, the experiment was divided into three successive phases: the first exposure phase, followed by the main experiment phase, and finally the last exposure phase.
%

\begin{figure}[t]
\newcommand\imgsize{4.00cm}
\centering
\includegraphics[height=\imgsize]{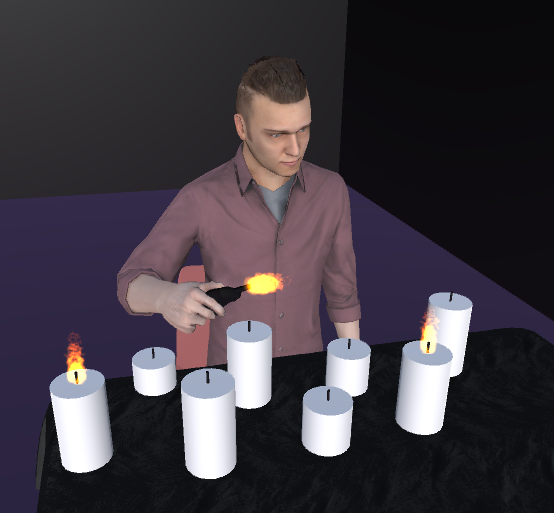}
\includegraphics[height=\imgsize]{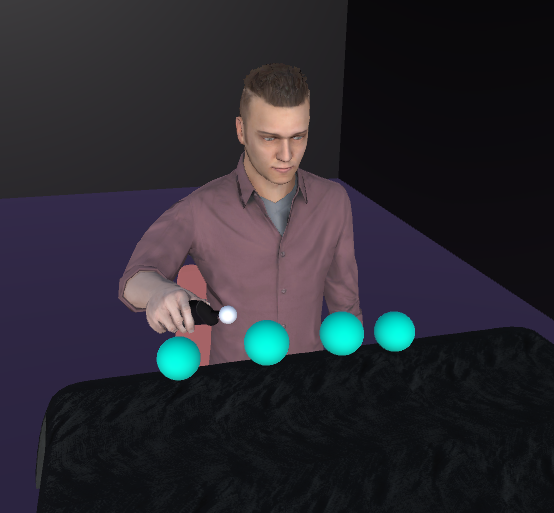}
\caption{Exposure phase in which participants were asked to light candles (left); main experiment in which participants were asked to touch some spheres with the tip of their controller (right). Images are shown from third-person perspective for illustrative reasons.}
\label{fig:experimentPhases3PPOV}
\end{figure}

\begin{figure}[t]
\newcommand\imgsize{2.90cm}
\centering
\includegraphics[height=\imgsize]{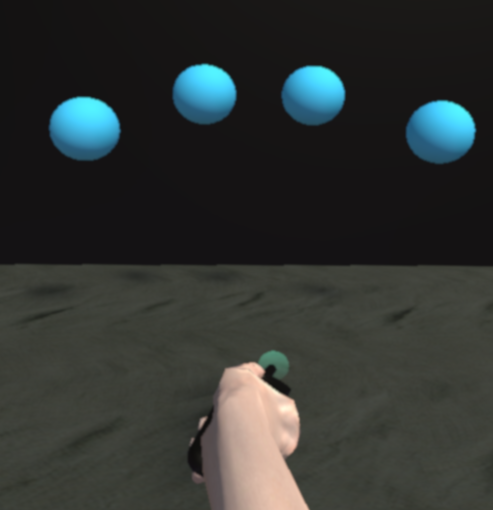}
\includegraphics[height=\imgsize]{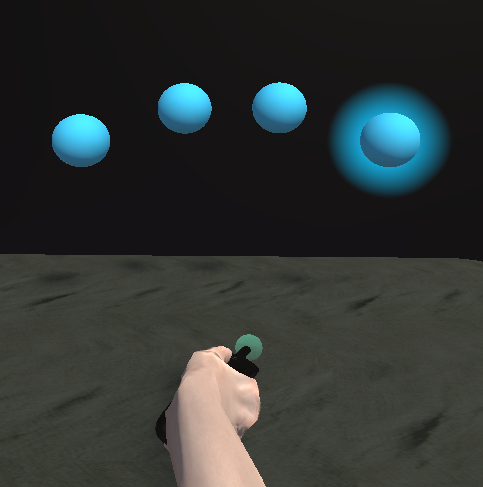}
\includegraphics[height=\imgsize]{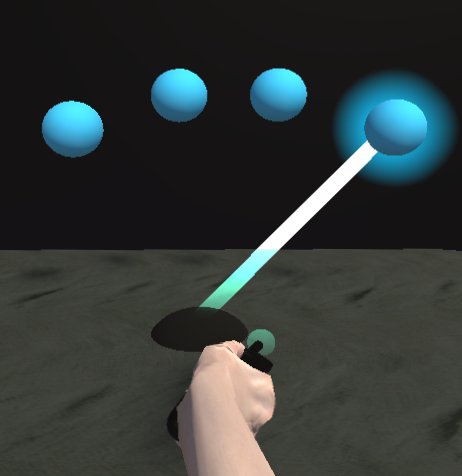}
\caption{The three tasks that the users were asked to perform. Free task: participants had to choose which sphere to touch (left). Target task: the sphere that the participants were to touch was highlighted (center). Trajectory task: the sphere to touch was highlighted and participants had to follow a path from the table to the sphere with the tip of their controller (right).}
\label{fig:SphereTasks}
\end{figure}

\begin{enumerate}
    \item[$\bullet$ First exposure phase:] The first exposure phase was conducted to allow the users to be accustomed to the shared body control and experimental environment (see Figure~\ref{fig:experimentPhases3PPOV}). Moreover, we took advantage of this phase to evaluate users' SoA and SoBO to assess their level of embodiment when possessing full (independent body) or half control (shared body) over the virtual avatar.
    \item[$\bullet$ Main experiment:] To explore the influence of the level of shared control toward the avatar on the SoA, five controlling weights were considered between 0\% and 100\%. In addition, to evaluate the influence of the freedom of movement and the intention toward an action on the SoA, three tasks were considered (Figure~\ref{fig:SphereTasks}).
    \item[$\bullet$ Last exposure phase:] This phase was conducted to evaluate \added{potential training effects} of the main experiment over agency and ownership ratings.
\end{enumerate}

\subsection{Participants}

Twenty-four male participants from the university campus participated in the experiment [age: $M=26 \pm$5 (SD)]; they were recruited from among both students and staff. They were all unaware with respect to the purpose of the experiment, had normal or corrected-to-normal vision, and gave written and informed consent. The study conformed to the declaration of Helsinki. The participants were paired with those they had never interacted with earlier. Among the participants, seven had no previous experience with VR, fourteen had limited previous experience, and three were familiar with VR. All participants were right-handed male Caucasians, to match the visual appearance of the virtual avatar as much as possible.
%


\subsection{Experimental Protocol}

\begin{figure}[t]
\centering
\includegraphics[width=\columnwidth]{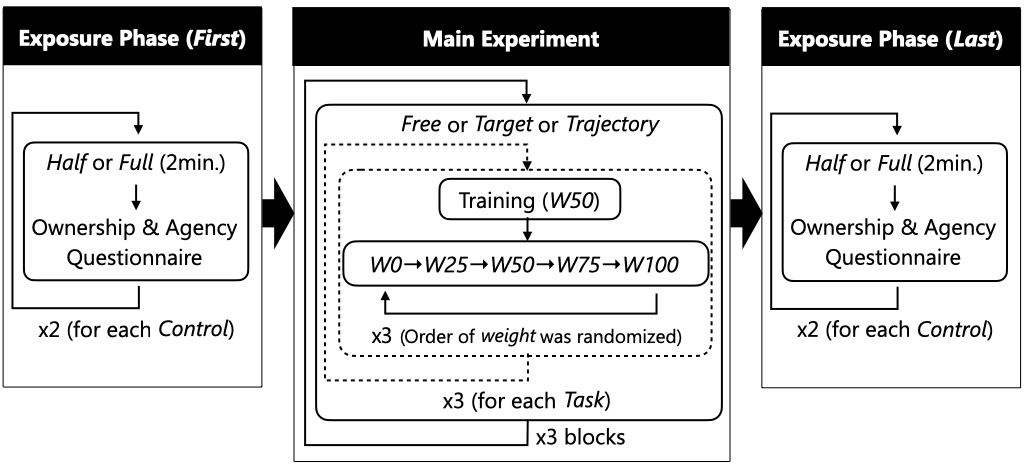}
\caption{Diagram of experimental flow.}
\label{fig:XPdiagram}
\end{figure}

The overall organization of the experiment is summarized in Figure~\ref{fig:XPdiagram} and is further described as follows.

Upon their arrival, participants read and signed the experiment consent form and filled in a demographic questionnaire. They also completed the IPC cognitive
test~\cite{LevensonIPC}. The internal score computed from this test was used later to measure LoC and explore its influence on the SoA. Then, they were briefed about the experiment through an explanatory video. They were explained that they would share a body and control over it with the other participant. After the explanation, they were instructed to sit on a chair in front of a table facing each other and wear an HMD to get immersed in the VE (Figure~\ref{fig:physicalSetUp}).

As previously explained, the experiment was divided into three phases, which the participants experienced in order: the first exposure phase, main experiment, and last exposure phase. In addition, while participants were immersed in the VE, they were instructed not to talk or interact with each other. As the tasks to perform only required motions of the right arm, we decided to focus on the right arm and did not animate the left arm. Participants were therefore asked to keep their left arm along their torso and not move it. After the experiment, they were instructed to remove their HMDs and provide general comments and feedback through a web form. The overall process took approximately 1 h.

\subsubsection{First and last exposure phases}

Participants started with the first exposure phase and finished with the last exposure phase, in which they were asked to light candles using a virtual lighter (Figure~\ref{fig:experimentPhases3PPOV}, left). Once participants had lit all their candles, the candles would extinguish, and the participants were asked to light them again. This task lasted for 2 min, and each phase was repeated twice (2 blocks): once with half of the avatar control for each participant, and once with full control over their own avatar. Each block would finish with an ownership and an agency questionnaire, which consisted of 11 items (Table~\ref{table:questionnaire}); the participants answered based on a scale ranging from 1 to 7 by pressing buttons on the controller in their hand. Each participant thus answered the questionnaires four times.


\subsubsection{Main experiment} \label{subsubsec:mainexperiment}

In the main experiment, the avatar was always shared and the weight of avatar control for a participant varied between 0\% and 100\% (respectively 100\% minus this weight for the other participant). We considered five weights between 0\% and 100\% to evaluate how differences in the degree of control would impact participants' SoA. Thus, we hypothesized that the SoA would be positively correlated with the degree of control.
%

Participants were asked to perform three tasks involving touching one virtual sphere among four spheres, with the extremity of virtual controller hold in the right hand. Four spheres were presented in front of the participants, all at equivalent distances from their right hand.
More precisely, by using the original 3D model of the HTC Vive controller, we attached a short rod with a small sphere on top; this tip collided with the sphere (Figure~\ref{fig:experimentPhases3PPOV}, right).

There were three types of tasks: free, target, and trajectory. The different tasks contrasted from each other with respect to the freedom of movement they allowed and whether both participants possessed the same prior knowledge of the same action to perform.
More precisely, in the free task, each participant was free to choose which sphere to touch (Figure~\ref{fig:SphereTasks}, left). In the target task, the sphere to touch was highlighted with a colored halo (see Figure~\ref{fig:SphereTasks}, center). Similarly, in the trajectory task, the sphere to touch was highlighted and the participants were asked to follow a path from the table to the highlighted sphere by using the tip of the controller; this task required more precision (see Figure~\ref{fig:SphereTasks}, right).

These three tasks were selected in line with the hypothesis that constraints in the movements and prior knowledge of the action to perform (i.e., the intention toward the action) both impact the SoA. In the free task, each participant was free to choose which sphere to touch (Figure~\ref{fig:SphereTasks}, left), under a condition where the movement of participants was not restricted and where the movement intention was not assuredly shared as participants might not decide to touch the same sphere. In the target task, the sphere to touch was highlighted with a colored halo (Figure~\ref{fig:SphereTasks}, center), under the condition that the movement was not restricted and the movement intention was shared as both participants focused on touching the same sphere. In the trajectory task, the sphere to touch was highlighted and participants were to follow a path from the table to the sphere by using the tip of the controller. This task required more precision (Figure~\ref{fig:SphereTasks}, right), and included both movement restriction and the shared intention, as participants had to follow a specific path to touch the same sphere.

These choices were driven by the demonstration of previous studies that SoA increases when participants have more action choices~\cite{Barlas2017}.
However, in our case, owing to changes in the level of control over the avatar, the more the participants had the choice of the action (in the free task compared to the target and trajectory tasks), the more the visuomotor discrepancies were expected between participants and avatar movements.
We thus supposed that the SoA would be higher for the target and trajectory tasks with smaller visuomotor discrepancies. Considering the results of Wegner et al.~\cite{Wegner2004}, we also expected that SoA would be higher in tasks where the intention of movement was shared (in target and trajectory tasks compared to free task).

In each task, participants performed 45 trials. For each trial, the participants started observing their own avatar over which they had full control. 
To ensure that both participants had the same initial position, they were asked to place their right hand holding the controller on the table on a specific virtual reference and remain on the initial reference. After 2 s, the four spheres were displayed in red with a message ``don't move yet''.
The message disappeared after 2 s, the spheres turned blue, and then the participants could perform the task. When a sphere (any of the four spheres for the free task, specified sphere in other tasks) was touched for 1 s by the tip of the controller of the shared avatar, the task was over and the following question was asked to both participants: ``On a scale ranging from 1 to 7, how much did you feel in control during this trial?''. As such, we followed the same protocol as that used by Jeunet et al.~\cite{Jeunet2018} to assess the SoA through a question that is easily understandable by participants and proved to relate to the judgment of agency. Participants provided a rating between 1 and 7 to validate their choice using the controller. When both participants had answered the question, they were asked to place their hand on the highlighted spot to start the next trial.

\subsection{Experimental Design}


\subsubsection{First and Last Exposure Phases}

\footnotesize
\begin{small} 

\begin{table}[t]
\begin{centering}
\small
\caption{Questionnaire used in the first and last exposure phases. Questions in italics are control questions regarding agency and ownership. }
\label{table:questionnaire}
\begin{tabular}{|c|L{6.5cm}|c|}
\hline
Variable & Question  \\ \hline
Agency & 1) The virtual arm moved just like I wanted to, as if it was obeying my will.  \\        
& 2) I felt as if I was controlling the movement of the virtual arm. \\ 
          & 3) \textit{I felt as if the virtual arm was controlling my will.} \\ 
          & 4) \textit{I felt as if the virtual arm was controlling my movements.} \\ 
          & 5) \textit{I felt as if the virtual arm had a will of its own.}  \\ 
          & 6) I felt as if I was causing the movement I saw. \\ \hline
Ownership  & 1) I felt as if I was looking at my own arm.  \\ 
           & 2) I felt as if the virtual arm was part of my body. \\ 
           & 3) I felt as if the virtual arm was my arm. \\ 
           & 4) \textit{I felt as if I had no longer a right arm, as if my right arm had disappeared.}  \\ 
           & 5) \textit{I felt as if the virtual arm was from someone else's body.}  \\ \hline
 \end{tabular}
\end{centering}
\end{table}

\end{small}
\normalsize

A within-subject design was set up for these experimental phases, where we considered two independent variables: \textit{control} and \textit{stage}. The main variable (\textit{control}) considered whether the participants were sharing the avatar, and possessed two levels: 1) participants sharing the avatar with 50\% control each (\textit{Half}) or 2) participants having full control over their own avatars (\textit{Full}). The \textit{stage} variable determined whether the task was completed in the first or last part of the experiment, and thus had two levels: \textit{First} and \textit{Last}. This part of the experiment was divided into two blocks corresponding to the two levels of the control condition. In both first and last exposure phases, whether participants would start with one block or the other was fully counterbalanced in the experiment.

The measured data (dependent variables) in a questionnaire were inspired from previous work~\cite{Botvinick1998,Longo2008,KALCKERT2014118}, where questions were divided in two groups: agency and ownership (Table~\ref{table:questionnaire}). For each question, participants were asked to
provide rating based on a 7-point Likert scale.

Based on previous works showing that asynchronous visual information in relation to participants' own movements affects both SoBO~\cite{Banakou2014, Kalckert2012, Ma2015} and SoA~\cite{Franck2001a, Farrer2008}, our main hypothesis was that participants would have lower agency and ownership when they had only half control than when they had full control of their avatar.

\subsubsection{Main Experiment}

We also adopted a within-subject design for the main part of the experiment, considering two independent variables: \textit{weight} and \textit{task}. The \textit{weight} variable determined the degree of control the participants had over the avatar and had five levels (\textit{W0, W25, W50, W75, and W100}). For each pair, \textit{weight} was inverted between participants, i.e., the sum of the controlling weights of the two participants was always 100\%. \textit{Task} corresponded to the three tasks included in the experiment (\textit{Free, Target}, and \textit{Trajectory}; see Section~\ref{subsubsec:mainexperiment} for details). The main experiment was divided into three blocks. To minimize the ordering effect, the orders of the blocks and tasks were counterbalanced following a Latin square design. 
Each iteration of \textit{Task} in one block comprised one training trial (with half control of the avatar) and three repetitions of the five trials (for the five levels of \textit{Weight}). The order of \textit{Weight} levels within the three repetitions was fully counterbalanced. Without considering the training trials, each participant performed 135 trials. Each trial lasted around 3 s.

The measured data (dependent variables) considered the performance and behavioral measurements. Regarding performance, we measured task-completion time, i.e., the time required to select the sphere after it turned blue (in seconds). Regarding behavioral measures, the motions (position and orientation per frame) of the participants' and shared avatar's controllers were recorded during the trials. Finally, the subjective FoC ratings for the question, ``How much did you feel in control?'' asked after each trial were rated on a 7-point Likert scale. Participants also reported general comments and feedback at the end of the experiment.

In summary, considering our experimental design, our main hypotheses are as follows. 

\begin{itemize}
\item[\textbf{H1}] When the degree of control (\textit{Weight}) decreases, the FoC ratings decrease. 
\item[\textbf{H2}] The FoC ratings will be higher for the tasks in which movements are more constrained (Trajectory~$>$~ Target~$>$~Free). 
 \item[\textbf{H3}] Participants with a higher Internal score of LoC experience higher FoC.

 
 
\end{itemize}

\section{Results}
\label{sec:results}

\begin{figure}[t]
\centering
\includegraphics[width=0.495\columnwidth]{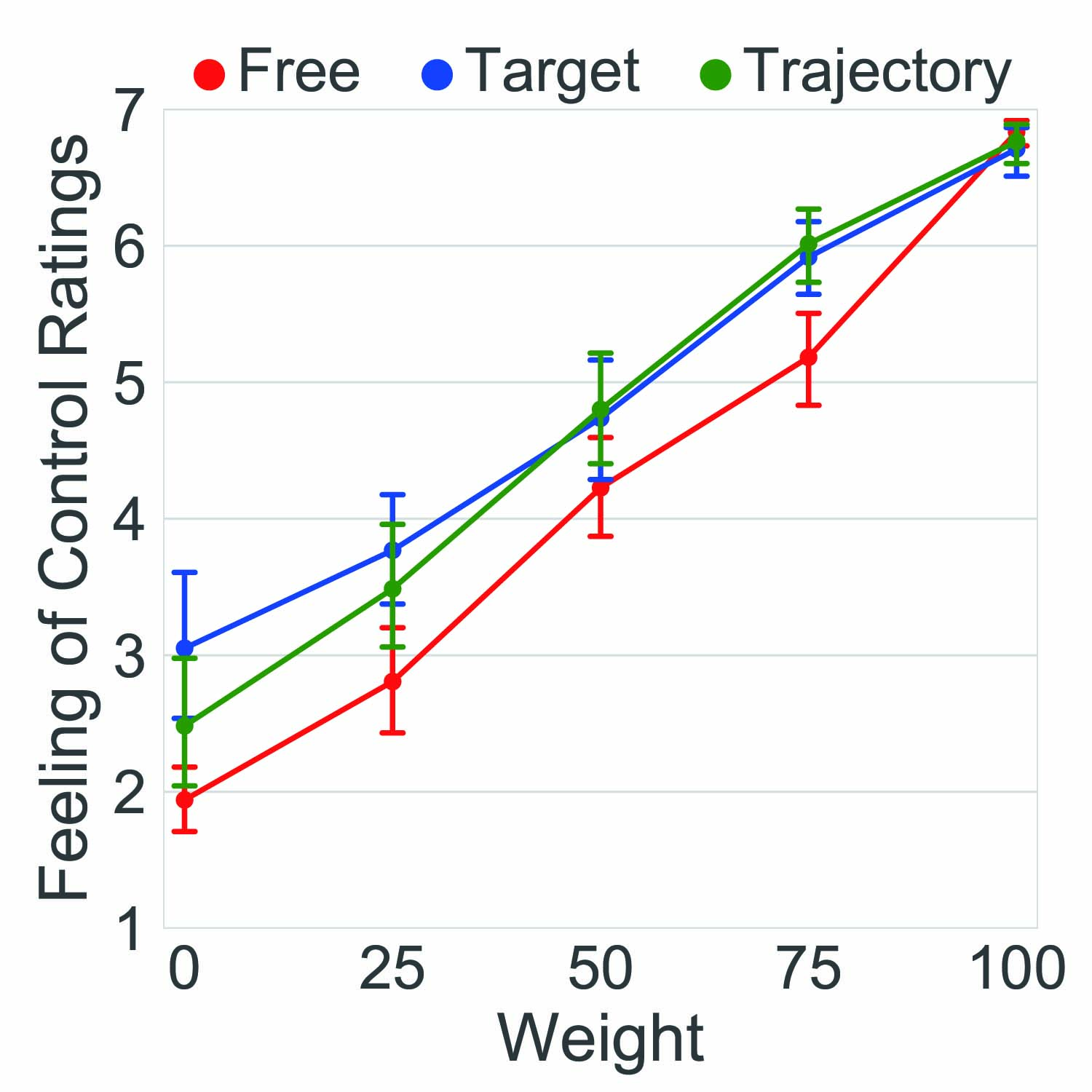}
\includegraphics[width=0.495\columnwidth]{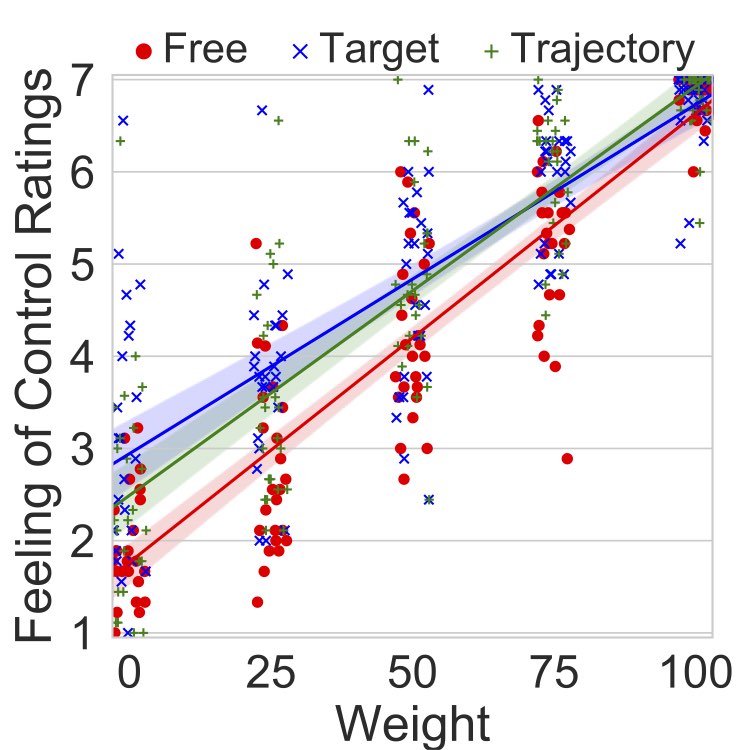}
\caption{
Left: Point plot of the mean subjective ratings of Feeling of Control (FoC) considering Weight of control and Task during main experiment. Right: Scatter plots with linear regression lines of FoC ratings on Weight for each task (\textit{Free}: $R^2$=0.83, \textit{Target}: $R^2$=0.66, \textit{Trajectory}: $R^2$=0.74). Error bars (left) and translucent bands (right) indicate 95\% CIs. A total of 10,000 bootstrap samples were used to estimate each 95\% CI.}
\label{fig:SoAmain}
\end{figure}

\subsection{Main Experiment}

Eleven trials out of all 3240 trials were excluded from the analysis after a visual inspection of the raw data revealed that either the task completion time, participant motion, or motion of the avatar exhibited abnormal values (values outside the range of three standard deviations from the mean).
ANOVA analyses were conducted when the normality assumption (Shapiro--Wilk's normality test) was not violated ($p>.05$). In particular, two-way ANOVA analyses with repeated measures were conducted, considering the within-group factors of Weight (5 levels: \textit{W0}, \textit{W25}, \textit{W50}, \textit{W75}, and \textit{W100}) and Task (3 levels: \textit{Free}, \textit{Target}, and \textit{Trajectory}).
When the sphericity assumption was violated (Mauchly’s sphericity test), the degrees of freedom were corrected using the Greenhouse--Geisser correction.
In addition, $\eta^2_p$ was provided for the quantitative comparison of effect sizes.
Finally, Tukey's post-hoc tests ($\alpha=.05$) were conducted to check the significance for pairwise comparisons of the parametric data.

When the normality assumption was violated (Shapiro--Wilk's normality test, $p<0.05$), Friedman test was conducted for each task independently followed by a post-hoc Wilcoxon-signed ranks test.
For multiple post-hoc comparisons, Holm correction was applied for the non-parametric data.
As for the correlation analyses, Pearson's r ($r$) was used for parametric data and Spearman's r ($r_s$) was used for non-parametric data.

\subsubsection{Feeling of Control (FoC)} \label{sec:FoC}

The two-way ANOVA analysis revealed a significant main effect of Task [\anovaETAbody{1.84}{42.37}{17.07}{<}{.001}{0.43}] and of Weight
 [\anovaETAbody{2.4}{55.15}{256.86}{<}{.001}{0.92}].
%
%
%
However, the two-way ANOVA also exhibited a significant interaction effect between Task and Weight [\anovaETAbody{5.22}{120.01}{6.30}{<}{.001}{0.22}].
%
%
First, Tukey's post-hoc tests indicated that, for all tasks, the FoC significantly decreased as the degree of control (Weight) decreased ($p<.001$ for all), which is further supported by the primary effects on Weight. Thus, this result supports~\textbf{[H1]}.
%
%
Next, when comparing the FoCs for each Weight level (see Figure \ref{fig:SoAmain} left), Tukey's post-hoc tests demonstrated that, for the \textit{W0} Weight, the FoC was significantly higher for the \textit{Target} task than for the other tasks (both $p<.05$).
%
%
%
In contrast, for the \textit{W25}, \textit{W50}, \textit{W75} levels of Weight, the FoC was significantly lower for the \textit{Free} task than for the other tasks (all $p<.05$).
%
%
Finally, for the \textit{W100} Weight, the post-hoc tests did not exhibit any significant difference.
Thus, these results only support \textbf{[H2]} partially, as although the \textit{Free} task (the less constrained task) consistently obtained the lowest FoC ratings (except for the \textit{W100}), this effect was not visible between \textit{Target} and \textit{Trajectory} tasks.

%
%

As the ANOVA analysis indicated that the strongest effects originated from the Weight factor, to further characterize the relationship between FoC and the Weight factor, a linear regression analysis was conducted across participants for each task (Figure \ref{fig:SoAmain} right).
The regression equations were

\begin{align*}
\textnormal{\textit{Free}: } y& = 0.0487x + 1.77 (R^2=0.83) \\
\textnormal{\textit{Target}: } y& = 0.0379x + 2.94 (R^2=0.65) \\
\textnormal{\textit{Trajectory}: } y& = 0.0444x + 2.49 (R^2=0.73).
\end{align*}

The regression equations exhibited linear positive correlations between the FoC and the Weight.
To determine whether the computed slopes differed significantly from 0, we computed the slope of each participant’s linear regression and conducted a t-test ($H_0$: Slope is equal to 0): (\textit{Free}: {\it t}(23)=35.665, {\it p} $<$ .001, 
\textit{Target}:  {\it t}(23)=13.219, {\it p} $<$ .001, 
\textit{Trajectory}: {\it t}(23)=16.622, {\it p} $<$ .001).
The results of the t-test indicated that the mean slopes all significantly differed from 0.
These results further support \textbf{[H1]}.
Section~\ref{sec:LoC} further analyzes the FoC ratings in correlation with the IPC scores.

\subsubsection{Task Completion Time}

\begin{figure}[t]
\centering
\includegraphics[height=0.42\columnwidth]{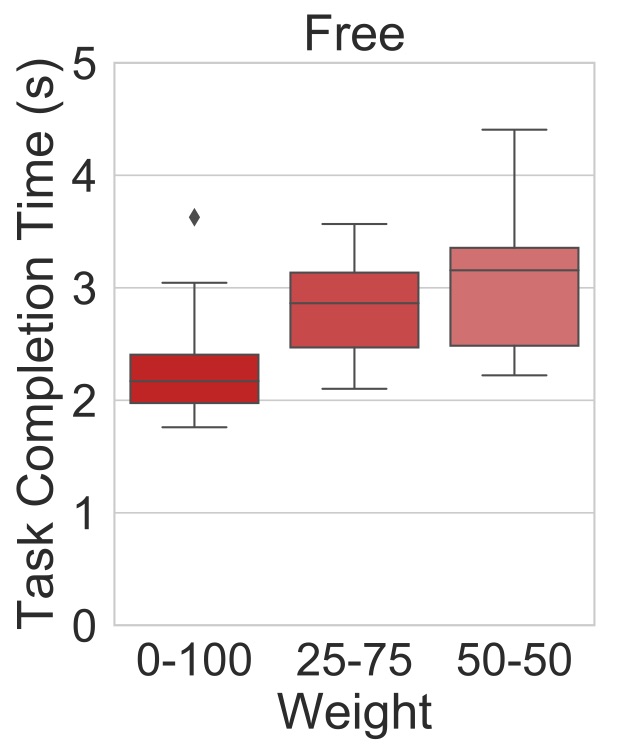}
\includegraphics[height=0.42\columnwidth]{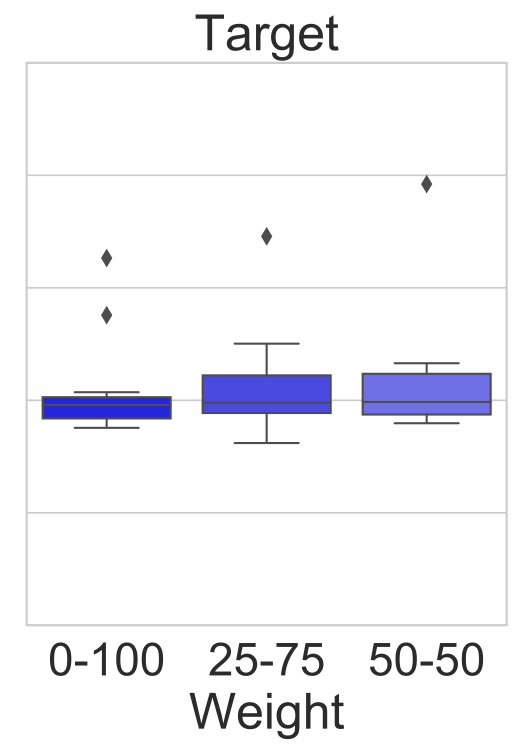}
\includegraphics[height=0.42\columnwidth]{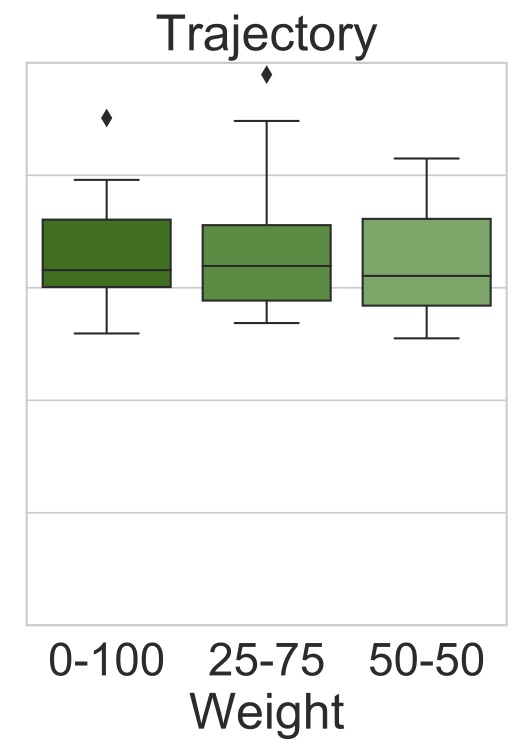}
\caption{Box plots of the task completion time considering the different Weight groups for each task.}
\label{fig:Time}
\end{figure}

Because the task completion time was dependent on the weights of the two participants (their sum adding to 100\%), for the task completion time analysis, the Weight group
factor had only three levels: \textit{W0-W100}, \textit{W25-W75}, and \textit{W50-W50} (see Figure~\ref{fig:Time}).
In addition, owing to the different natures of each task (aimed movement, path following task), we did not assess the differences among Tasks for the task completion time.
Therefore, we conducted three Friedman tests considering Weight group as a factor, one for each task.
The Friedman tests exhibited significant differences among the task completion times of the Weight groups only for
the \textit{Free} task ($\chi^2$=14, {\it p} $<$ .001), and no significant differences were found for the \textit{Target} task ($\chi^2$=0.17, {\it p} $=$ .92) or
the \textit{Trajectory} task: ($\chi^2$=3.5, {\it p} $=$ .17).
Post-hoc pairwise comparisons indicated that for the \textit{Free} task, the task completion time was significantly smaller in the \textit{W0-W100} condition
(W0-W100 $<$ W25-W75: {\it Z}=-2.81, {\it p}$<$.01, 
W0-W100 $<$ W50-W50: {\it Z}=-3.30, {\it p}$<$.01). 
No significant differences were found between \textit{W25-W75} and \textit{W50-W50} ({\it Z}=-1.68, {\it p} $=$ .092).
%


%
\subsubsection{Motion Data}

\begin{figure}[t]
\centering
\includegraphics[height=0.40\columnwidth]{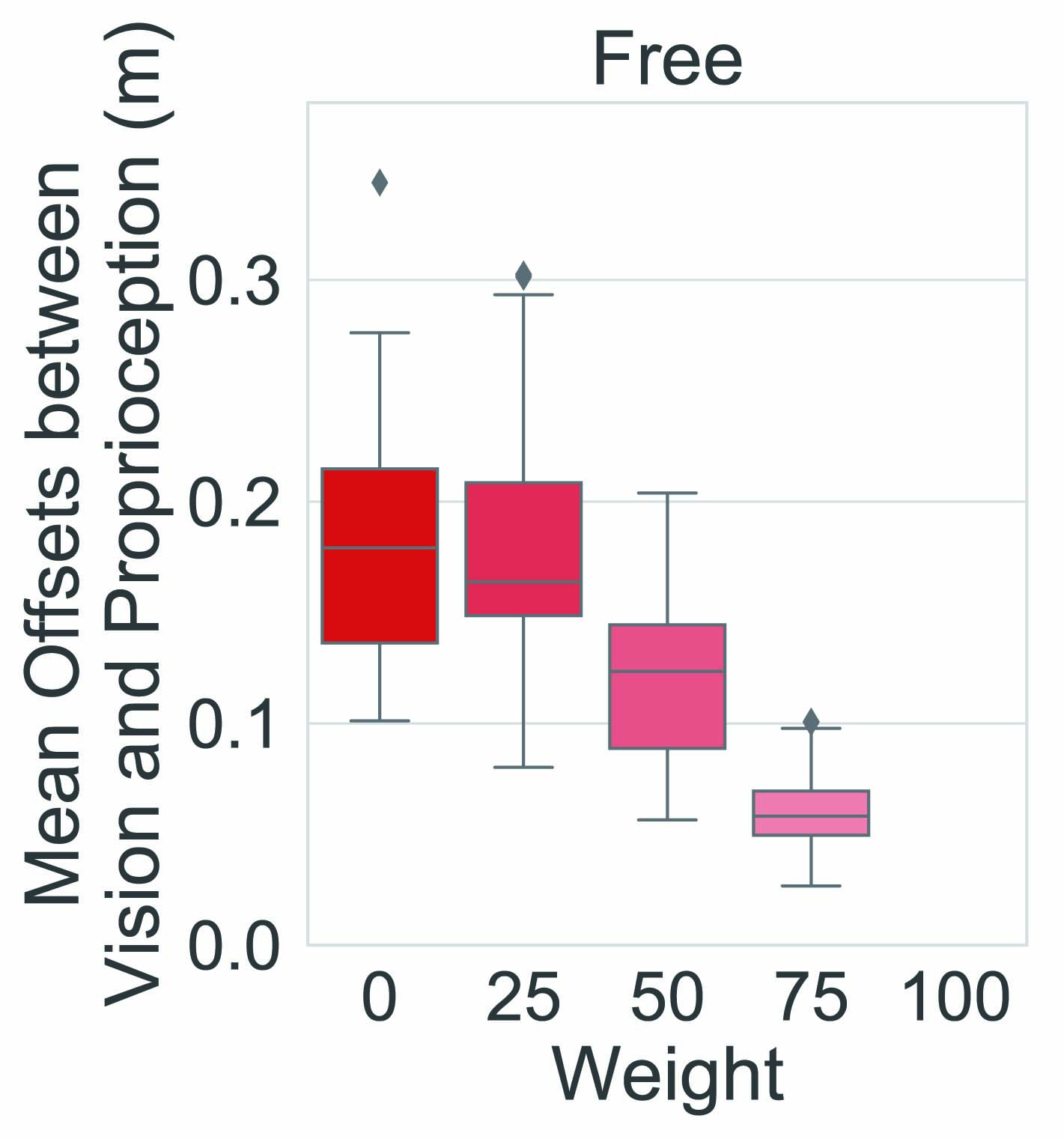}
\includegraphics[height=0.40\columnwidth]{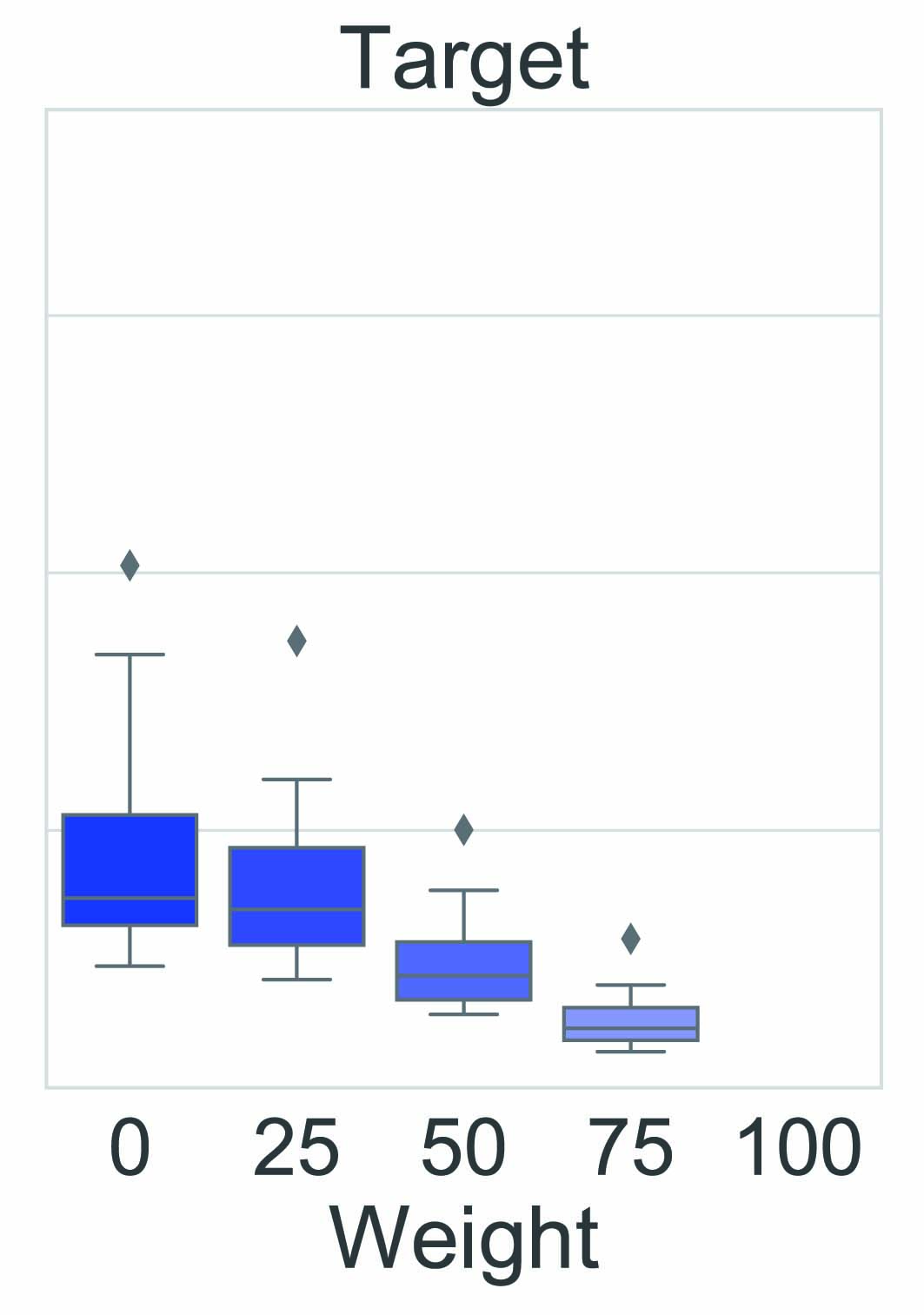}
\includegraphics[height=0.40\columnwidth]{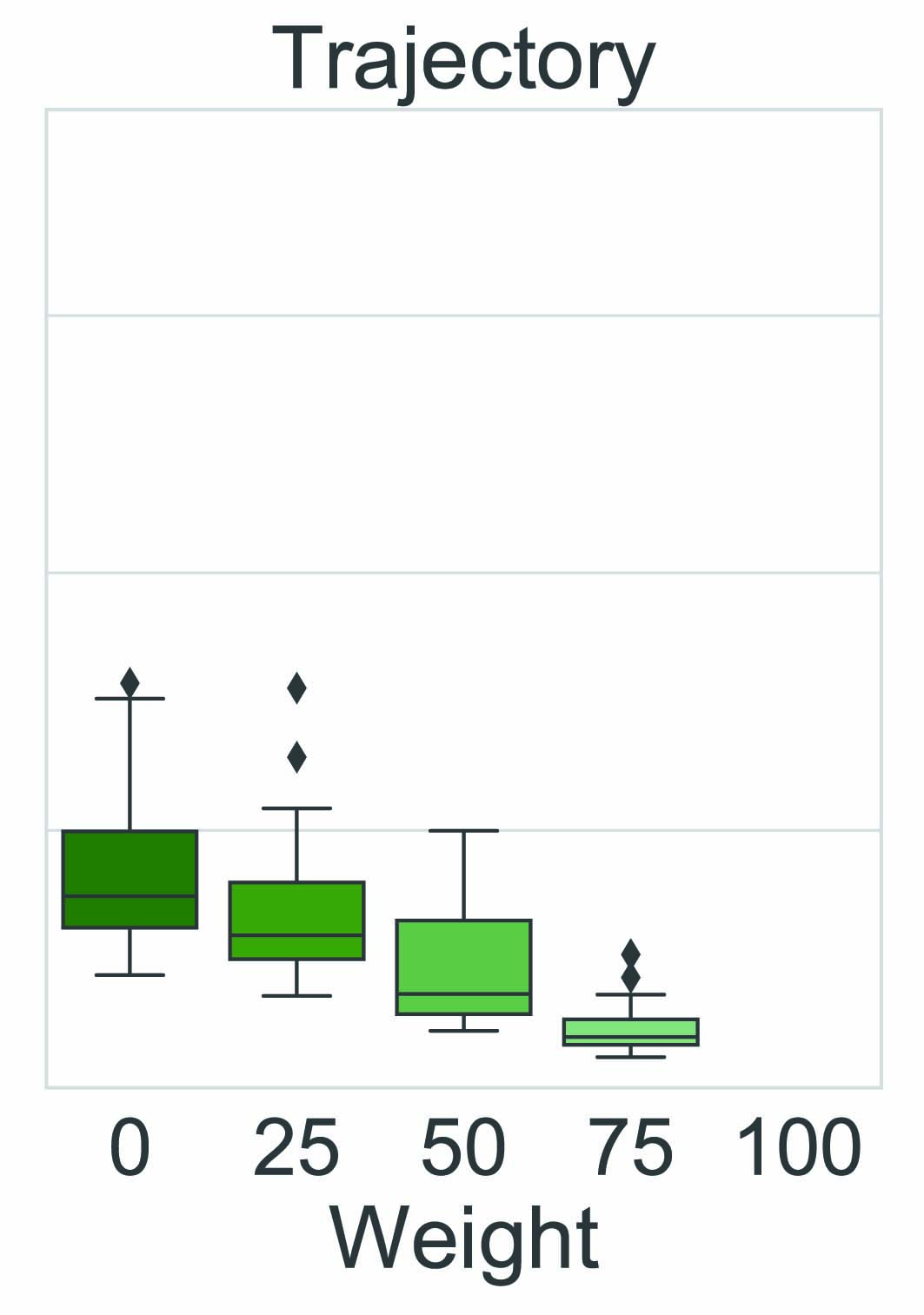}
\caption{Box plots of the mean offsets between the positions of the avatar's hand and the participant's actual hand considering the Weight for each Task.}
\label{fig:OffsetBoxPlot}
\end{figure}

\begin{figure}[t]
\centering
\includegraphics[width=\columnwidth]{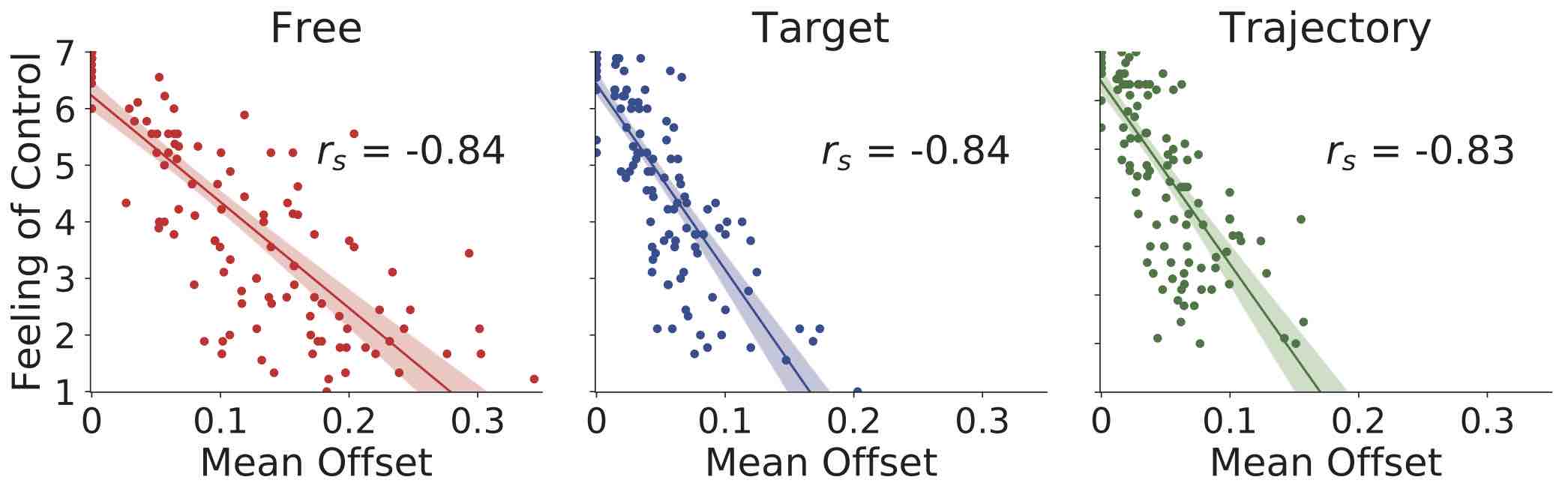}
\caption{Scatter plots with linear regression lines of FoC ratings on mean offsets between the positions of the avatar's hand and  the participant's actual hand for each Task.
Translucent bands indicate 95\% CIs.
}
\label{fig:CorrDist}
\end{figure}

The offsets (Euclidean distance) between the positions of the participant’s and avatar’s hands were calculated for each frame and then averaged for each trial (see Figure~\ref{fig:OffsetBoxPlot}).
This value provided a rough estimate of the overall visuo-motor discrepancies for each trial.
We excluded the \textit{W100} condition from the analysis as the discrepancy was 0 regardless of the Task (condition with full control).
The residuals did not follow a normal distribution; thus, Friedman tests were considered. In addition, the analysis considered each Task independently.
Friedman tests exhibited significant differences of the mean offsets among Weights for all Tasks: 
(\textit{Free}: $\chi^2$=56.75, {\it p} $<$ .001, 
\textit{Target}: $\chi^2$=67.25, {\it p} $<$ .001, 
\textit{Trajectory}: $\chi^2$=61.85, {\it p} $<$ .001).
Post-hoc pairwise comparisons indicated that the mean offsets were significantly smaller when the Weight was larger for all comparisons in all Tasks ({\it p}$<$.001 all) except for the comparison between offsets in the \textit{W0} and \textit{W25} conditions for the \textit{Free} task. 
%


An additional correlation analysis was performed to assess the link between the mean offset across all weights and the perceived FoC.
%
The correlation analysis revealed that the offsets were negatively correlated with FoC for all Tasks:
\textit{Free}: {\it $r_s$}=-0.84, {\it p}$<$.001, 
\textit{Target}: {\it $r_s$}=-0.84, {\it p}$<$.001, 
\textit{Trajectory}: {\it $r_s$}=-0.83, {\it p}$<$.001) (See Figure \ref{fig:CorrDist}). 

%
Moreover, to check if the mean offsets would vary between tasks, another analysis was performed for each weight. Friedman tests revealed significant differences among the mean offsets of Tasks for W0 ($\chi^2$=28.58, {\it p} $<$ .001), W25 ($\chi^2$=32.33, {\it p} $<$ .001), W50 ($\chi^2$=37.33, {\it p} $<$ .001), and W75 ($\chi^2$=32.33, {\it p} $<$ .001).
Post-hoc pairwise comparisons showed that for W0, W25, W50, and W75 the mean offsets were significantly higher for \textit{Free} compared to \textit{Target} and \textit{Trajectory} (both {\it p}$<$.001).

\begin{figure}[t]
\centering
\includegraphics[width=\columnwidth]{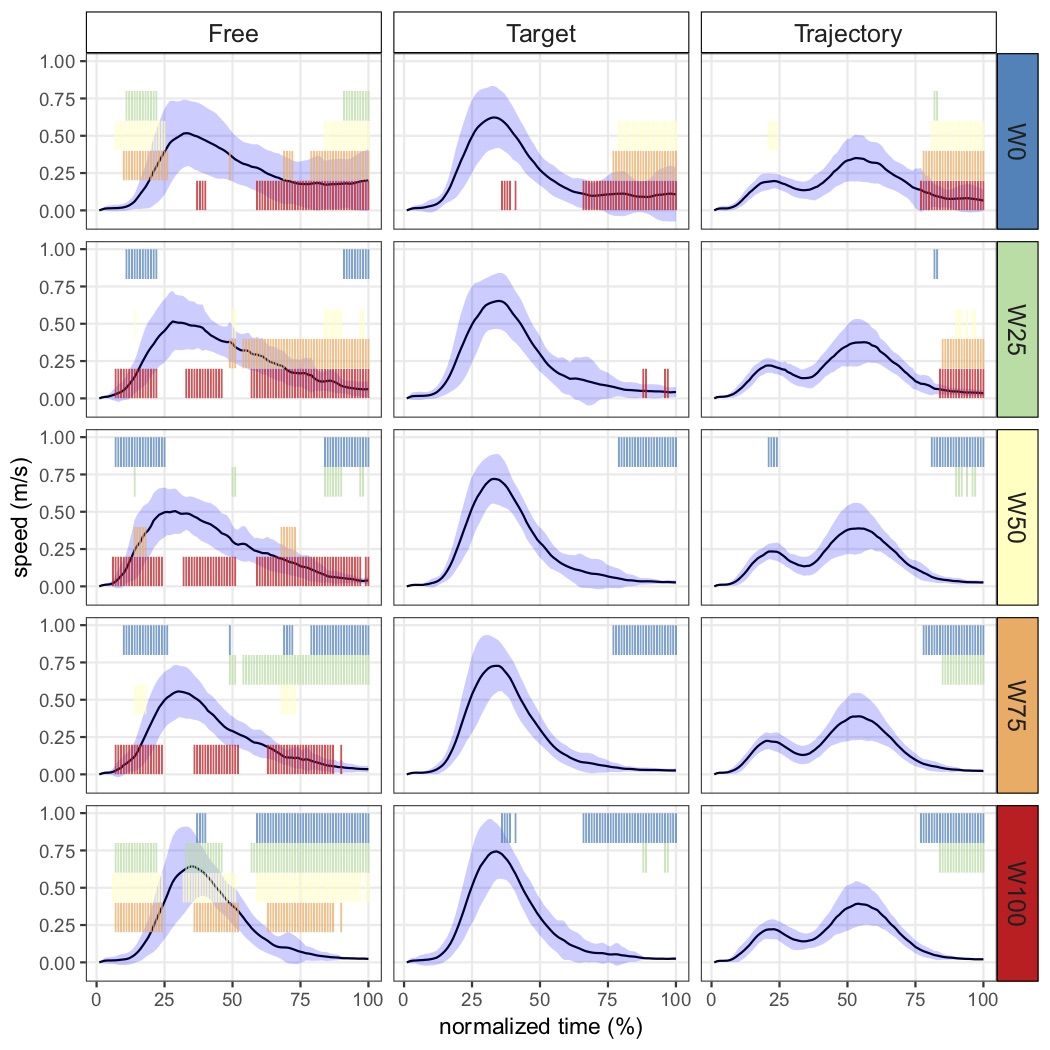}
\caption{Averaged Speed profiles between all participants for each Weight and Task, normalized in time and re-sampled at 100 intervals. Colored segments were placed at intervals in which significant differences (Friedman and Wilcoxon pairwise comparison tests) were found. Colors are associated to a specific Weight, from the lightest (lower Weight) to the darkest (higher Weight).}
\label{fig:AverageSpeedProfile}
\end{figure}

Finally, to gain some insight regarding the global behavior of users during each trial, speed profiles were computed for each participant per Weight and Task for each trial. Speed profiles were normalized in time by resampling the values at 100 intervals between the start (time 0\%) and end of the trial (time 100\%). We then computed the mean and standard deviation of the speed profiles between all participants as reported in Figure~\ref{fig:AverageSpeedProfile}.
To compare the speed profiles for each Task and for each interval, we conducted a Friedman test considering Weight as a factor. Tasks were not compared among each other as the nature of each Task was different.
Among those intervals, post-hoc pairwise comparisons (Wilcoxon signed-rank tests) were performed to find pairwise differences among different Weights.
The results of pairwise comparisons are also summarized in Figure~\ref{fig:AverageSpeedProfile}, in which each Weight is denoted by a color; lighter colors are associated with lower Weights and vice-versa, and colored segments are placed at the intervals in which significant differences were found. 
Thus, the presence of a colored segment indicates that a significant difference ({\it p} $<$ .05) was found between the current interval and the corresponding interval of the color-coded condition.
This result allows us to highlight the tasks in which changes in the control induced differences in participant behavior. For example, for \textit{Target} and \textit{Trajectory} tasks, the Weight seems to only have a visible impact at the end of the motion, in particular for \textit{W0} and \textit{W25}, whereas more discrepancies were found for the \textit{Free} task.



\subsection{Personality Traits} \label{sec:LoC}

\begin{figure}[t]
\centering
\includegraphics[width=0.9\columnwidth]{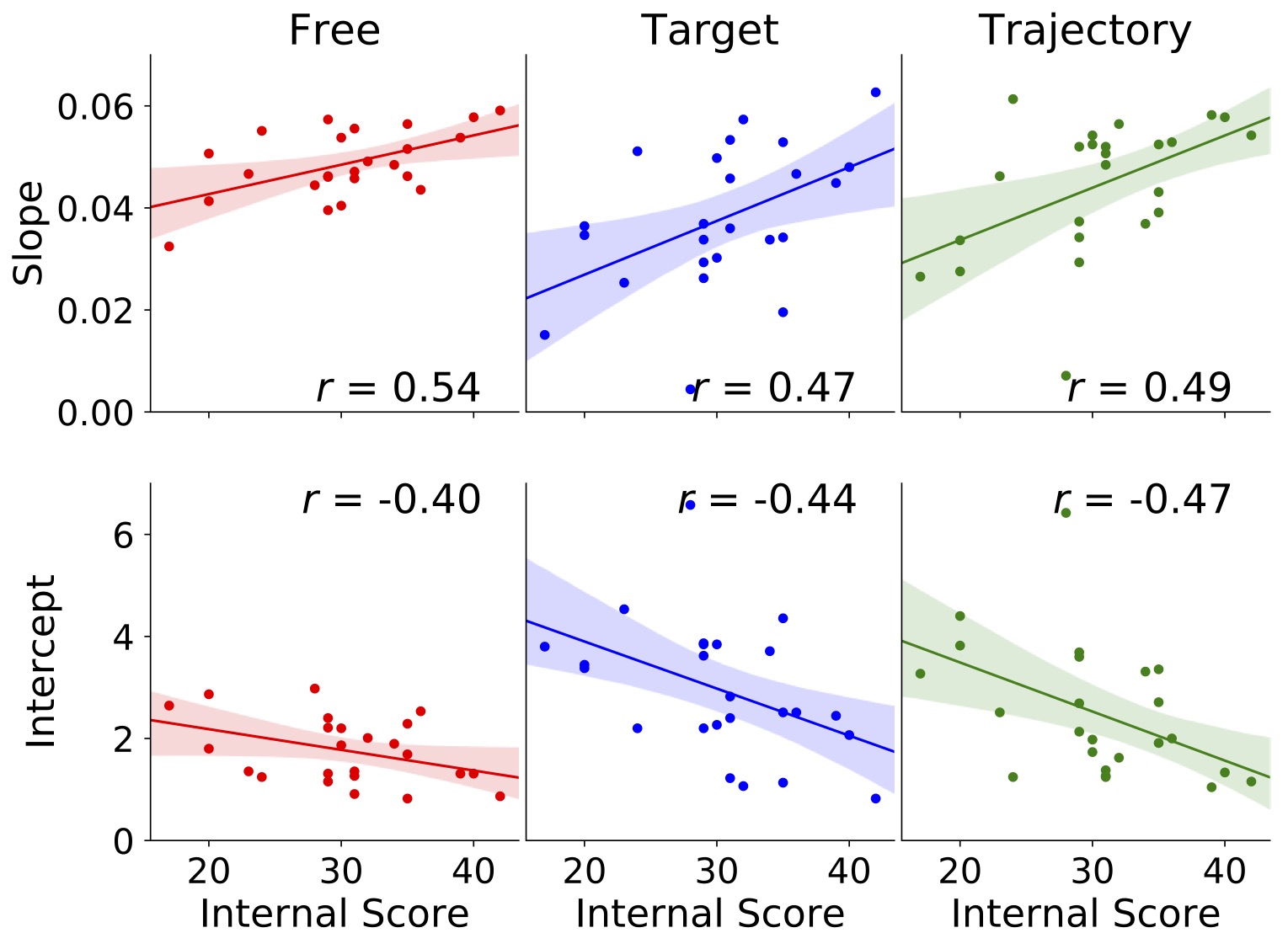}
\caption{Scatter plots with linear regression lines between the internal score of the IPC test and the regression coefficient terms obtained between the FoC ratings and Weight for each participant (slope (top) and intercept (down)).
Translucent bands indicate 95\% CIs.
10,000 bootstrap samples were used to estimate each 95\% CI.}
\label{fig:IPC}
\end{figure}

According to the responses of the IPC test, each participant obtained three scores (from 0 to 48), one for each dimension of the IPC test (i.e. Internal, Powerful Others, and Chance). Each score was calculated by adding the responses of the eight items for each dimension and a constant of 24.
Similar to previous studies, only the Internal dimension was assessed~\cite{Jeunet2018}, as it was found to be the dimension that was more related to the FoC.
A high rating on the Internal score indicates that the subject has a strong internal Locus of Control (i.e., they believe that events in their life derive primarily from their own actions).

First, to verify whether participants with higher internal score of IPC tended to experience higher FoC 
when they had full control (W100), we conducted a correlation analysis between the internal scores and the mean FoC scores in the W100 condition for each task.
As a result, no significant correlation was found between the internal score and the FoC: \textit{Free} ({\it $r_s$}=0.23, {\it p} $=$ .29), \textit{Target} ({\it $r_s$}=0.33, {\it p} $=$ .11), \textit{Trajectory} ({\it $r_s$}=0.25, {\it p} $=$ .23). 
%
This result might be explained by a ceiling effect of very high values of FoC in the W100 condition. This result does not support~\textbf{[H3]}.

In contrast to previous studies, the modulation of the participant's control was quantified by the Weight parameter. This enables us to analyze the correlation of the internal component of the IPC with the FoC in a wider range of FoC values.
First, as already detailed in Subsection~\ref{sec:FoC}, we computed the correlation between the Weight and the FoC for each participant. The intercept coefficient could be considered as the FoC ``baseline,'' while the slope could be related to the ``sensitivity'' to changes in the participant's control. In other words, the slope provides information on how much the change in the participant's control influences the FoC, and the intercept provides a lower bound for the FoC. In practice, in our scenario, both parameters are strongly correlated because there is a strong ceiling effect for the FoC at W100. 
Thus, we computed the regression equations of FoC on Weight for each participant and performed correlation analyses of both the slopes and intercepts with the participants’ score of the Internal dimension of the Locus of Control (from the IPC test). The results show a positive correlation between the slope and the Internal dimension for each Task (\textit{Free}: {\it r}=0.54, {\it p}$<$.01, \textit{Target}: {\it r}=0.47, {\it p}$<$.05, \textit{Trajectory}: {\it r}=0.49, {\it p}$<$.05, see Figure \ref{fig:IPC} up), as well as the negative correlation between the intercept and the Internal dimension for \textit{Target} and \textit{Trajectory} tasks (\textit{Target}: {\it r}=-0.44, {\it p}$<$.05, \textit{Trajectory}: {\it r}=-0.47 {\it p}$<$.05), and a marginally significant effect for the \textit{Free} task ({\it r}=-0.40, {\it p}=.05) (See Figure \ref{fig:IPC} down). 

%
These results seem to suggest that participants with a higher Internal score were more sensitive to changes in the avatar control as they had lower intercept values and higher slope values.

\subsection{First and Last Exposure Phases}


\begin{figure}[t]
\centering
\includegraphics[height=0.37\columnwidth]{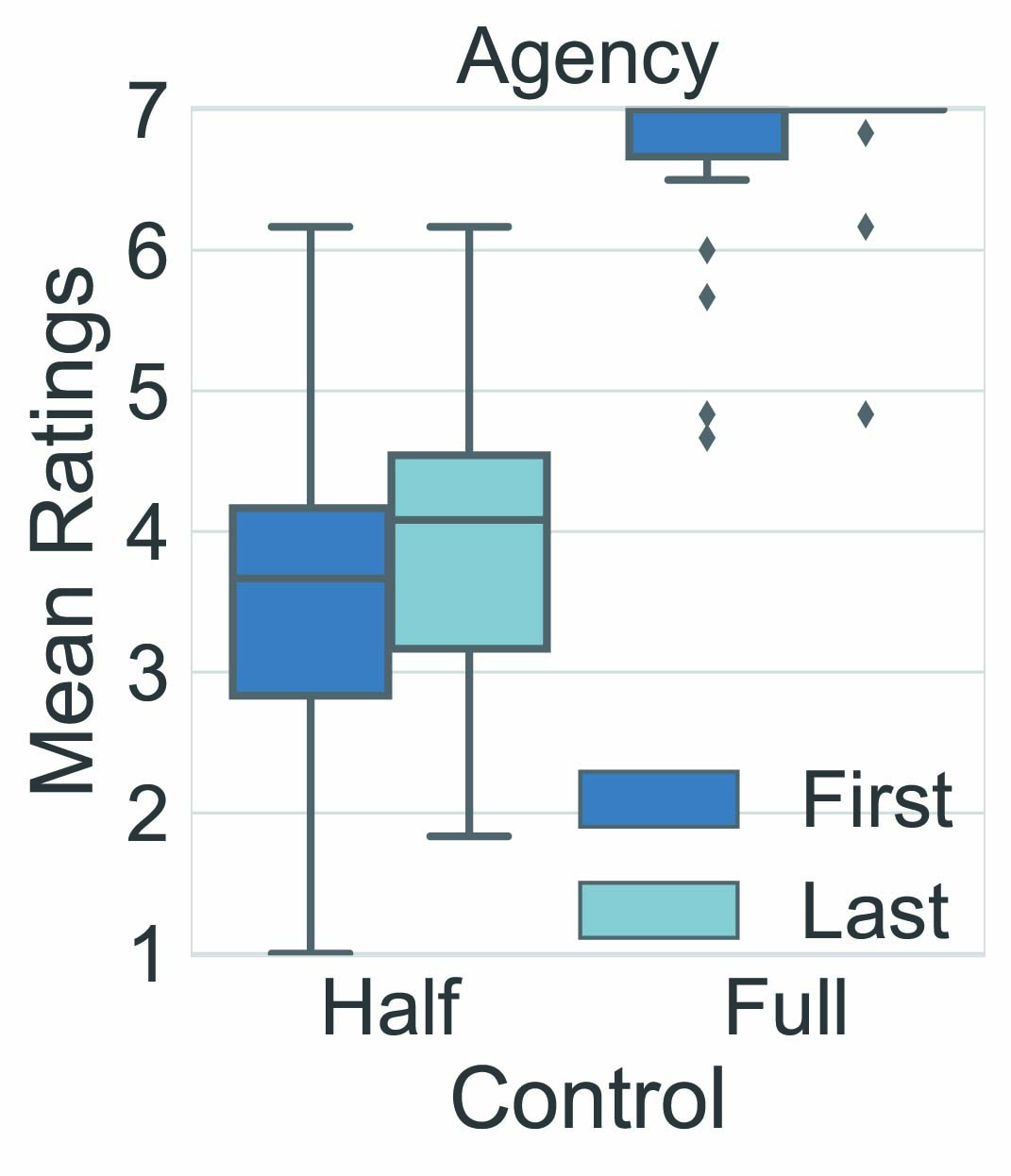}
\includegraphics[height=0.37\columnwidth]{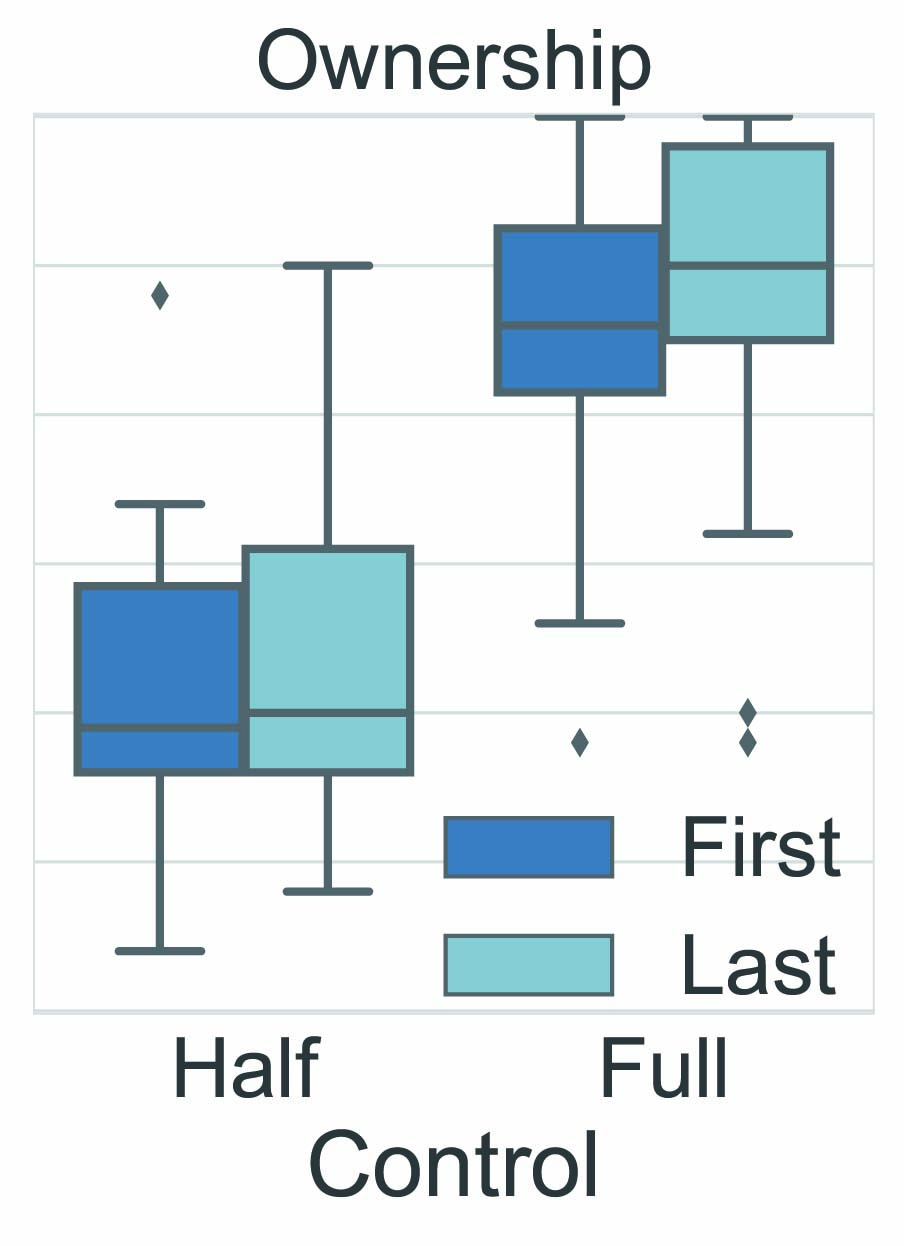}
\includegraphics[height=0.37\columnwidth]{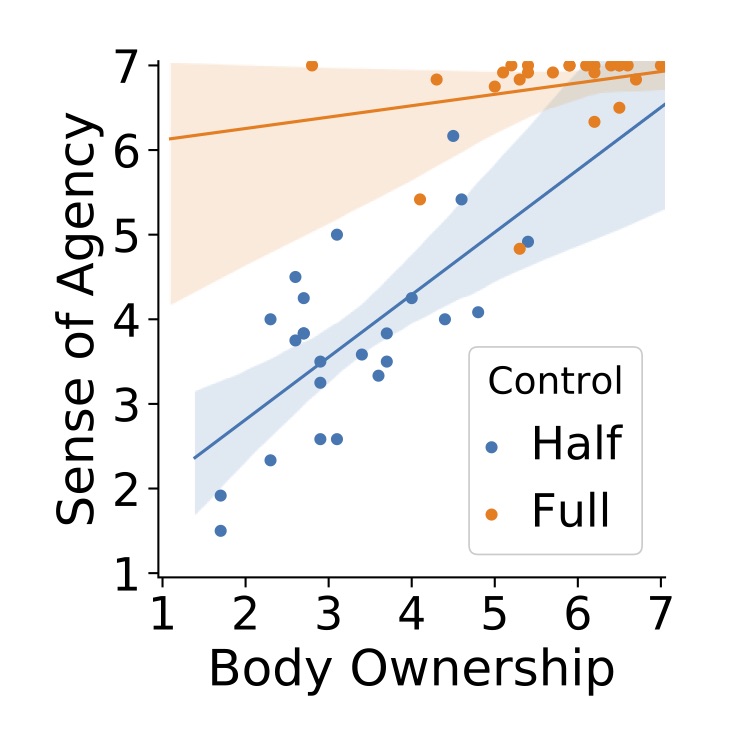}
\caption{Box plots of mean ratings of agency (Left) and ownership (Center) obtained in the questionnaires in \textit{First} and \textit{Last} exposure phases. Right: Scatter plots with linear regression lines of agency ratings on ownership ratings. Translucent bands indicate 95\% CIs. 10,000 bootstrap samples were used to estimate each 95\% CI.}
\label{fig:prepost}
\end{figure}

The agency and ownership ratings for the \textit{First} and \textit{Last} exposure phases were aggregated and averaged (control item answers were inverted) to compute one agency and one body ownership score per participant.
Owing to the non-parametric nature of the data and the need of testing interaction effects, we applied an aligned rank transform (ART) on the data. This procedure enables the use of ANOVA to analyze the interaction effects with non-parametric data~\cite{Wobbrock2011}. 
Two-way repeated measures ANOVAs with the within-subjects factors Control (2 levels: \textit{Half} and \textit{Full}) and Stage (2 levels: \textit{First} and \textit{Last}) were performed for both agency and ownership scores.
Regarding the agency scores, the ANOVA revealed a significant main effect of 
Weight [\anovaETAbody{1}{23}{198.41}{<}{.001}{0.90}] 
and Stage [\anovaETAbody{1}{23}{19.22}{<}{.001}{0.46}] 
(Figure~\ref{fig:prepost} Left).
In addition, the Weight $\times$ Stage interaction effect was significant [\anovaETAbody{1}{23}{5.17}{<}{.05}{0.18}].
Thus, we only report the post-hoc tests for the interaction effect.
%
First, post-hoc pairwise comparisons using the Wilcoxon signed-rank test (Holm corrected) showed that in both \textit{First} and \textit{Last} phases the agency
scores were significantly higher in the \textit{Full} condition than in the \textit{Half} condition (\textit{First}: {\it Z}=-5.29, {\it p}$<$.001, \textit{Last}: {\it Z}=-5.29, {\it p}$<$.001). 
%
Second, in both \textit{Full} and \textit{Half} conditions, the agency scores were higher in the \textit{Last} than \textit{First} phases (Full: {\it Z}=-2.58, {\it p<.05}, Half: {\it Z}=-2.09, {\it p<.05}).

Regarding the ownership scores, the ANOVA showed a main effect of Weight [\anovaETAbody{1}{23}{84.96}{<}{.001}{0.79}] %
and a marginally significant main effect of Stage [\anovaETAbody{1}{23}{3.78}{=}{.06}{0.14}] %
(See Figure~\ref{fig:prepost} Center).
The Weight $\times$ Stage interaction effect was not significant 
[\anovaETAbody{1}{23}{0.68}{=}{.42}{0.03}].
Similar to the agency ratings, the ownership scores were significantly higher for the \textit{Full} condition.
In addition, we conducted a correlation analysis between the agency and the ownership scores for each participant, showing that ownership was positively correlated with agency in the \textit{Half} condition ({\it $r_s$}=0.54, {\it p}$<$.01), but not in the \textit{Full} condition ({\it $r_s$}=0.31, {\it p} $=$ 0.14) (See Figure~\ref{fig:prepost} Right).

\section{Discussion}
\label{sec:discussion}

In this section, we discuss how the results can be interpreted in terms of SoA, which is measured by subjective judgments of FoC over the participants' actions. We also provide additional insights regarding the Locus of Control and the relation between SoA and SoBO. 

\subsection{Main Results}


The SoA results show that changes in the degree of control clearly influenced the SoA, which validated \textbf{[H1]}. More precisely, the FoC ratings, which were treated as an explicit measure of the SoA according to previous studies~\cite{Wegner2004, Linser2007}, increased linearly with the increase in the degree of control for all three tasks.
This result can be explained by the fact that the higher the degree of control is, the closer the visual feedback of the avatar hand is to the actual hand position of the participant, thereby reducing visual mismatch between the movements of the avatar and the participants' actual movements. As stated by Farrer et al.~\cite{farrer:hal-00655447}, our ability to recognize SoA from the visual cues of movement tend to decrease in case of mismatch between visual feedback and actual movement, i.e., when there are visuo-proprioceptive discrepancies, which could justify the correlation observed between the SoA and the degree of control. The participants' feedback is also in line with this interpretation, as they expressed their disturbance when their arm was controlled out of their will: ``\textit{It was confusing when the hand was going in the direction I intended it to go but the speed did not totally match my movements}''. These results can also be explained by the phenomenon of ``body semantic violation'' introduced by Padrao et al.~\cite{PADRAO2016147}. In our case, it refers to the fact that the agency illusion will break when the discrepancy between feedback and intended motion become too important.

Another interesting result reveals that when participants had no control over the avatar (W0), the SoA was higher for the Target task than for the Free and Trajectory tasks. While we hypothesized that the nature of the task could influence the perceived SoA, the tasks differed in two main aspects. The first difference relates to whether participants shared an intention toward the action to perform. In the Target and Trajectory tasks, the sphere to be touched was indicated, meaning that both participants shared the same intention of action: touching the same sphere. On the contrary, in the Free task, participants could have different spheres to touch in mind; this sometimes resulted in a difference between the intention, the sphere a participant wanted to touch, and the resulting action, the sphere finally touched by the shared avatar.
According to Wegner et al.~\cite{Wegner1999}, SoA arises if (1) an intention precedes an observed action (priority), (2) the intention is compatible with this action (consistency), and (3) the intention is the most likely cause of this action (exclusivity). In the Target and Trajectory tasks, the three principles of priority, consistency, and exclusivity are more likely to be respected as participants share the same intention. Independently of their degree of control, the controller of the shared avatar will therefore reach the targeted sphere. This would support why SoA ratings where higher in the Target and Trajectory tasks when participants had no control over the shared avatar. 
%
%
%
The second difference was in the visual difference between participants and avatar hand positions (See Figure~\ref{fig:OffsetBoxPlot}) depending on the task. Indeed, results showed for example that visuo-motor and visuo-proprioceptive discrepancies
%
were lower in the Target task compared to Free when participants had no control. This can be because in the Target task, participants have the indication of which sphere to touch, resulting globally in the same movement toward the target sphere. Following the statements of Farrer et al.~\cite{Farrer2008} that visuo-motor discrepancies tend to decrease the SoA, this could explain why the SoA was higher in Target than in the Free task where participants had no control at all. However, these results only partially support \textbf{[H2]}.

Furthermore, a surprising result is that in the Target and Trajectory tasks, participants tended to overestimate their SoA, feeling some SoA despite the absence of control. From the analysis of speed profiles, we observed that major differences between control weights were found in the Free task, whereas only some differences were observed in the Target and Trajectory tasks, mostly at the end of trials. This seems to show that participants tended to have similar reaching behaviors regardless of their degree of control in the tasks where the goal was shared.
Other authors also observed that the SoA was affected when the avatar's and the participant's speed of movement differed~\cite{Kokkinara2015}, but not with spatial shift of movement without speed alteration.
These results could explain why participants tended to overestimate their SoA in the Target and Trajectory tasks, as we can see that even with no or very low control, participants still performed the task in a similar manner, therefore minimizing spatio-temporal discrepancies. 

We also observed during the experiment that some participants reported a pure illusion of the control: ``\textit{Sometimes, when the task was accomplished in an excellent manner, I wondered if it was actually me who had moved the arm }''. It is known how high-level contextual information (whether participants believe that the outcome is either triggered by themselves or by somebody else) can influence intentional binding~\cite{Desantis2011}, referring to the implicit measure of the conscious experience of SoA~\cite{Moore2012}. Depending on whether participants were more or less aware of their degree of control over the avatar may have affected their SoA. Furthermore, another feedback particularly illustrates potential future studies: ``\textit{I had the impression that sometimes no one controlled my movement and that I was actually watching a video}''. Indeed, sharing the control of the avatar with an autonomous virtual agent instead of another person would be an interesting topic to explore, in line with other studies which explored the influence of human and computer co-actors over the SoA in joint actions~\cite{Obhi2011}. In particular, they showed that SoA for self-generated actions was inhibited when the participants knew that a computer was the co-actor of the action, which would be interesting to explore in the context of our co-embodiment setup.

\subsection{SoA and Personality Traits}

%
According to the results of the correlation analyses between the slope or intercept of FoC and the internal dimension of the locus of control (Figure~\ref{fig:IPC}), the intercept of the regression of FoC scores on the weight factor was negatively correlated with the Internal scores, especially when participants had little or no control over the virtual body, which does not validate \textbf{H3}. More precisely, participants with a high Internal score tend to have their feeling of agency be more impacted by changes in the level of control.

In previous studies, the Internal score was observed to be positively correlated with participants’ SoA when participants were immersed in a VE and embodied in their own virtual avatar over which they had full control~\cite{Jeunet2018}. Our results do not support those findings, probably due to the ceiling effect we observed on SoA when participants had full control. However, we herein investigated the influence of the Locus of Control one-step further, exploring the influence of the Internal score on the SoA when participants did not have full control over their avatar. 
We found that participants with a high Internal score tend to have their SoA more impacted by changes in their degree of control of the avatar. People with a high Internal score are known to attribute the consequences to themselves rather than to chance or other more powerful entities and tend to believe that they have personal control over performance and rewards. However, such a definition does not commonly consider body movements. Given the little amount of previous work linking LoC and SoA, the results from such analyses should thus be treated with considerable caution. On the one hand, our results seem to suggest that people who tend to attribute consequences to themselves are possibly more aware of their own actions and thus notice more when they do not have control.
On the contrary, people with a high Internal score might attribute events, movements included, to themselves and then attribute the movements of the avatar they did not cause to themselves. We would thus expect from participants to experience a high SoA even with no control over the shared avatar. While our results are in contradiction with this hypothesis, it would be in agreement with Desantis et al.’s study~\cite{Desantis2011} wherein they showed that when participants believe that they have control over the environment, intentional binding, an implicit measure of the SoA, is stronger. However, in our analysis, we only tried to correlate the Internal score with FoC, an explicit measure of the SoA. As previous findings do not always agree on whether implicit and explicit measures of agency relate to the same thing \cite{10.1371/journal.pone.0110118}, it would be interesting to also consider correlating implicit measures of the SoA with the Internal score.
Therefore, our results on the influence of the Internal score of the Locus of Control over the SoA demonstrate the need for further investigation on the topic.


\subsection{Sense of Embodiment}

Results from the agency and ownership questionnaires in the first and last exposure phases showed that having only half the control of an avatar significantly decreased both SoBO and SoA compared to when they fully controlled an avatar (Figure~\ref{fig:prepost} Left and Center).
Such results are in line with numerous previous studies showing that asynchronous visual information with reference to participants' own movements
eliminates both SoBO~\cite{Banakou2014, Kalckert2012, Ma2015} and SoA~\cite{Franck2001a, Farrer2008}.
In addition, our results showed that agency and ownership scores were positively correlated when each participant had half of the control of the avatar, whereas no correlation was found when they had full control over their own avatar (see Figure~\ref{fig:prepost}, right). 
As for the relationship between SoBO and SoA, some studies indicate that both experiences can partially double dissociate~\cite{Kalckert2012, Sato2005, Braun2014, Tsakiris2010} while some others suggest that they may strengthen each other if they co-occur~\cite{Longo2008, Tsakiris2006, Banakou2014, Dummer2009} (For review, see~\cite{Braun2018}).
While we observed a ceiling effect of the agency scores when participants had full control, the
positive correlation found in the half condition indicates a close relationship between SoA and SoBO.
Furthermore, the variability of participants' responses suggest that the subjective experience of being embodied in a shared avatar vary strongly among individuals.

Considering such positive correlations, the induction of the stronger SoBO over the virtual body can be considered to make SoA stronger and vice versa.
Indeed, Kokkinara et al.~\cite{Kokkinara2015} observed that illusory SoA occurred despite the distortion of movements being larger than the detection thresholds of discrepancies found in previous studies. They also remarked that their results might be due to the full-body ownership illusion.
In our study, Figure~\ref{fig:CorrDist} indicates that in the Free task, participants felt more than half control when the distance between participant's and avatar's controller positions were below 0.1 m on average.
As SoBO is known to be affected by top--down factors such as the congruence of the structural and morphological features between one's own and virtual bodies~\cite{Kilteni2015}, making the features more congruent might therefore induce a stronger SoA. It is also considered to increase the detection threshold of visuo-motor discrepancies. In VR, some studies have exploited such visuo-motor discrepancies to enhance passive haptics or improve manipulations by changing the mapping of movements from the physical to the virtual space~\cite{Lecuyer2000, Azmandian2016, Kohli2012}. The interplay between SoBO and SoA is a subject of psychological interests, but seeking to reduce the detection threshold of visuo-motor discrepancies by strengthening SoBO might also be useful to VR applications.


In addition, there has been some evidence showing the dynamic relationship between self-attribution 
and sensorimotor systems.
In Nielsen's study~\cite{Nielsen1963}, participants experienced the illusory SoA and attributed the experimenter's hand in a mirror to their own while drawing a straight line.
In particular, when the experimenter distorted their movement so that he/she drew a curved line, participants still attributed the movement to themselves and moved in the opposition to the experimenter's movement to compensate for the error between the predicted and actual movements.
This means that as long as they attributed a movement to themselves, they tried to control it.
Asai~\cite{Asai2015} also reported that illusory self-attribution of fake movements might coordinate sensory input and motor output. Conversely, when participants became aware of the uncontrollability of the cursor, the illusory self-attribution was also dismissed.
In our experiment, the degree of control was different for each trial. Therefore, participants could not fully adapt to it.
However, in case of a constant degree of control, participants might feel a stronger SoA since visuo-motor adaptation might enable participants to predict the avatar's movements.
Investigation of the adaptation process of co-embodiment would therefore be necessary to further understand how to elicit higher SoA for future applications.


\subsection{SoA in Joint Action}

We perform joint actions together with others in our daily lives, e.g., carrying heavy things, and admirably coordinate our plans and actions to achieve our joint goal.
Indeed, in such cases, individuals build up a shared motor plan, incorporating others' actions into their own motor system during a joint action~\cite{Obhi2011}.
In joint actions, there is therefore an automatic formation of a new agentic identity (a 'we' identity)~\cite{Obhi2011}, and we feel the sense of us.

In the virtual co-embodiment situation where two individuals jointly control one avatar, as mutually coordinated actions of self and other produce the united movements, individuals might therefore also feel a sense of us.
In our experiment, we found it particularly surprising that participants were able to immediately coordinate their actions to the joint goal even with the completely novel way of interacting and the lack of verbal communication. 

Nevertheless, according to participants' feedback, this collaborative behavior was not shared between all participants and some of them even tended to get the feeling of competing while performing the task: ``\textit{I felt in competition especially for the free task}'', ``\textit{I sometimes felt in competition when we both had control and wanted to go on different spheres}''.
We also observed that the time to complete the task was higher when the control was equally shared between participants compared to when one participant had more control than the other in the Free task. Such differences could be caused by the adoption of ``leader/follower'' behaviours when one participant has more control that the other; however, further investigation would be necessary to explore such a hypothesis.
%
Overall, research on virtual co-embodiment could therefore contribute to studies of joint action that investigate the mechanisms of how individuals coordinate their actions online, which is the essential capacity of humans as social beings.

\subsection{Future Work}

\added{Despite the interesting insights gained from our experiment, we believe that there are still other aspects that would require further research.
}

\added{First, our study focused on a particular virtual co-embodiment experience, namely two users sharing an avatar to accomplish simple tasks with different degrees of control. The results showed that users were able to perform the tasks and their SoA positively correlated with their degree of control. Additionally, previous knowledge of actions to be performed significantly increased their SoA.
However, owing to the inter-relation of sharing the avatar and the actual degree of control, clearly quantifying the effects of each is difficult at this stage. Thus, the actual effect of being embodied in the same virtual avatar with someone else remains unclear. Does the mere fact of knowing that you share your avatar with someone else have an impact on the perception over the avatar? This is still an unanswered question that would require additional experiments, e.g., a virtual co-embodiment scenario in which a user shares the avatar with an autonomous agent.}


Second, the proposed control scheme demonstrated that a partial degree of control can still elicit an SoA over a shared virtual body and that the motor actions performed in such a context resemble the ones performed with full control of the virtual body. 
Our implementation was meant to evaluate a novel concept, for which we tested one of the potential shared-control schemes. For example, as the shared control of the avatar head was particularly problematic, we decided that each user would keep full control of the avatar head as sharing its control might require unwanted changes at the user's viewpoint. Such situations could lead the user to be prone to motion sickness. 
However, in situations where users are allowed to move freely around, a more complex scheme would therefore be required as the overall shared posture might be different than the users' own posture.
%
This would therefore require exploring more complex control schemes, techniques for switching control schemes depending on the situations and objectives, or even supporting more people embodied in the same avatar.
%
Moreover, even at the level of controlling individual body parts, different control schemes can be considered. In our implementation, we averaged the positions and orientations of the controllers, but other methods could, for instance, explore splitting the control of different body parts or taking control depending on a certain movement threshold. 

Third, virtual co-embodiment has a variety of potential applications such as remote training or entertainment.
In a manner similar to our method, Yang and Kim's ``Just Follow Me''~\cite{Yang2002} method visualizes the motion of the trainer as a ghost, superimposed on the avatar of the trainee in the virtual environment. A similar method was also proposed in augmented and mixed realities for remote guidance and collaboration~\cite{7563559, 10.1007/978-3-642-40483-2_5}.
In contrast, a system based on the principle of virtual co-embodiment could allow trainers to control a trainee's movements to different degrees depending on the training needs and allow them to interact with each other through body movements while sharing the same experience.
The results of our study showed that even when participants had no control over the avatar, they overestimated their FoC when the situation constrained the movements and indicated a shared goal.
It suggests that in the training situation, the trainee could feel SoA over their body even when the body is fully controlled by the trainer.
In addition, training could be made more effective by changing the degree of control depending on the learning phase, which in turn would require designing efficient and intuitive ways to adapt the degree of control to the situation.
Moreover, it would be interesting to compare the cognitive load inferred by our system with the one felt in an approach similar to the ``Just Follow Me'' method~\cite{Yang2002}, searching if one method is more susceptible to increase the cognitive load of the trainee while learning through an application.
This will also open new opportunities to explore how mismatching the actual and announced degrees of control influences the user's SoA, e.g., by
telling both users that they have a 75\% control even though they actually have 50\% control each.
Furthermore, another potential application of virtual co-embodiment could be the tele-operation of one robot by two experts at a time, as for instance the co-manipulation of a medical robot by two surgeons. In such a scenario, we may imagine experts taking alternatively more or less control over the avatar in order to actuate the robot, giving them the possibility of making ``pauses'' in the manipulation, while maintaining a first-person point of view in the avatar in order to keep following the procedure easily. Such applications could also be extended and relevant for tele-operations in asymmetric telepresence systems, as the one developed by Steed et al.~\cite{6353427}, where several users might be immersed in the same environment with different capabilities of interacting.
Overall, considering new means of making users efficiently collaborate in future applications, e.g., through the use of verbal interactions, visual cues, and interaction design, will be important.

Nevertheless, it must be emphasized that the results of this study were obtained only for male participants from the university campus (students and staff). Given that recent evidence suggests that interactions and collaboration between persons can be influenced by gender diversity (e.g., in teams~\cite{Bear2011}, in pedestrian interactions~\cite{vanBasten2009}), gender might have influenced the results of our study, particularly in terms of whether the participants adopted collaborative or competitive strategies. It would be valuable to replicate our study with participants of more diverse gender and attributes.


%

%
Lastly, as virtual co-embodiment is a merging experience with someone else, it has the possibility to produce cognitive effects on users.
Indeed, shared bodily experiences such as the enfacement illusion (i.e., self-other face-perception modification by synchronous multisensory stimulation)~\cite{Mazzurega2011a, Tsakiris2008, Sforza2010} are known to produce both perceptual and social binding.
A stranger stimulated in synchrony was judged as more similar, physically and in terms of personality, and as closer to the self~\cite{Mazzurega2011a, Tajadura-Jimenez2012a}.
In addition, enfacement was positively correlated with the participant's empathic traits and with the physical attractiveness that the participants attributed to their partners~\cite{Sforza2010}. 
In this sense, co-embodiment could be used as a tool for psychological investigations of the ``self''.
%
%



\section{Conclusion}
\label{sec:conclusion}

\added{In this paper, we introduced the concept of ``virtual co-embodiment'', a situation that enables a user and another entity (e.g., another user, robot, autonomous agent) to be embodied in the same virtual avatar. 
In addition, we described an experiment that examined the influence of the degree of control of an avatar shared with another person on one’s own SoA, as well as the influence of the predictability of avatar movements.}
%
Our results indicated that participants succeeded frequently in estimating their actual level of control over the shared avatar. Interestingly, they tended to overestimate their feeling of control when the visual feedback of the avatar's movements was closer to their actual movements, as well as when they had prior knowledge of the action to be performed.
In addition, our results showed that participants performed similar motions regardless of their level of control. 
Finally, our results reveal that the internal dimension of the locus of control is negatively correlated with the participants' perceived FoC.

%

Taken together, these findings not only corroborate and extend previous studies, but they also pave the way for further applications in the field of VR-based training and collaborative tele-operation applications in which users would be able to share their virtual body.

\ifCLASSOPTIONcompsoc
  \section*{Acknowledgments}
\else
  \section*{Acknowledgment}
\fi

We wish to thank 
the participants who took part in our experiment. This work was sponsored by the Region Bretagne, the Inria IPL Avatar project, and the JSPS Overseas Challenge Program for Young Researchers.

\ifCLASSOPTIONcaptionsoff
  \newpage
\fi



\bibliographystyle{IEEEtran}
%


\bibliography{biblio,coembodiment}

\begin{thebibliography}{10}
\providecommand{\url}[1]{#1}
\csname url@samestyle\endcsname
\providecommand{\newblock}{\relax}
\providecommand{\bibinfo}[2]{#2}
\providecommand{\BIBentrySTDinterwordspacing}{\spaceskip=0pt\relax}
\providecommand{\BIBentryALTinterwordstretchfactor}{4}
\providecommand{\BIBentryALTinterwordspacing}{\spaceskip=\fontdimen2\font plus
\BIBentryALTinterwordstretchfactor\fontdimen3\font minus
  \fontdimen4\font\relax}
\providecommand{\BIBforeignlanguage}[2]{{%
\expandafter\ifx\csname l@#1\endcsname\relax
\typeout{** WARNING: IEEEtran.bst: No hyphenation pattern has been}%
\typeout{** loaded for the language `#1'. Using the pattern for}%
\typeout{** the default language instead.}%
\else
\language=\csname l@#1\endcsname
\fi
#2}}
\providecommand{\BIBdecl}{\relax}
\BIBdecl

\bibitem{Kilteni2012}
K.~Kilteni, R.~Groten, and M.~Slater, ``{The Sense of Embodiment in Virtual
  Reality},'' \emph{Presence Teleoperators Virtual Environ.}, vol.~21, no.~4,
  pp. 373--387, 2012.

\bibitem{Hoyet2016}
L.~Hoyet, F.~Argelaguet, C.~Nicole, and A.~L{\'{e}}cuyer, ``{“Wow! I Have Six
  Fingers!”: Would You Accept Structural Changes of Your Hand in VR?}''
  \emph{Front. Robot. AI}, vol.~3, no. May, pp. 1--12, 2016.

\bibitem{8260949}
T.~C. Peck, M.~Doan, K.~A. Bourne, and J.~J. Good, ``The effect of gender
  body-swap illusions on working memory and stereotype threat,'' \emph{IEEE
  Transactions on Visualization and Computer Graphics}, vol.~24, no.~4, pp.
  1604--1612, 2018.

\bibitem{Latoschik2017}
M.~E. Latoschik, D.~Roth, D.~Gall, J.~Achenbach, T.~Waltemate, and M.~Botsch,
  ``The effect of avatar realism in immersive social virtual realities,'' in
  \emph{Proceedings of the 23rd ACM Symposium on Virtual Reality Software and
  Technology}, 2017, pp. 39:1--39:10.

\bibitem{10.1371/journal.pone.0189078}
Y.~Pan and A.~Steed, ``The impact of self-avatars on trust and collaboration in
  shared virtual environments,'' \emph{PLOS ONE}, vol.~12, no.~12, p. e0189078,
  2017.

\bibitem{Yang2002}
U.~Yang and G.~J. Kim, ``{Implementation and evaluation of "just follow me": An
  immersive, VR-based, motion-training system},'' \emph{Presence Teleoperators
  Virtual Environ.}, vol.~11, no.~3, pp. 304--323, 2002.

\bibitem{Kawasaki2010}
H.~Kawasaki, H.~Iizuka, S.~Okamoto, H.~Ando, and T.~Maeda, ``{Collaboration and
  skill transmission by first-person perspective view sharing system},'' in
  \emph{Proc. - IEEE Int. Work. Robot Hum. Interact. Commun.}, 2010, pp.
  125--131.

\bibitem{Luria:2019:RCE:3322276.3322340}
M.~Luria, S.~Reig, X.~Z. Tan, A.~Steinfeld, J.~Forlizzi, and J.~Zimmerman,
  ``Re-embodiment and co-embodiment: Exploration of social presence for robots
  and conversational agents,'' in \emph{Proceedings of the 2019 on Designing
  Interactive Systems Conference}.\hskip 1em plus 0.5em minus 0.4em\relax ACM,
  2019, pp. 633--644.

\bibitem{Kilteni2015}
K.~Kilteni, A.~Maselli, K.~Kording, and M.~Slater, ``Over my fake body: body
  ownership illusions for studying the multisensory basis of own-body
  perception,'' \emph{Front. Hum. Neur.}, vol.~9, 2015.

\bibitem{Longo2008}
M.~R. Longo, F.~Sch{\"{u}}{\"{u}}r, M.~P. Kammers, M.~Tsakiris, and P.~Haggard,
  ``{What is embodiment? A psychometric approach},'' \emph{Cognition}, vol.
  107, no.~3, pp. 978--998, 2008.

\bibitem{Wegner2004}
D.~M. Wegner, B.~Sparrow, and L.~Winerman, ``{Vicarious agency: Experiencing
  control over the movements of others},'' \emph{J. Pers. Soc. Psychol.},
  vol.~86, no.~6, pp. 838--848, 2004.

\bibitem{gomezhal-01228890}
G.~Gomez, C.~Plasson, F.~Elisei, F.~No{\"e}l, and G.~Bailly, ``{Qualitative
  assessment of an immersive teleoperation environment for collaborative
  professional activities in a ''beaming'' experiment},'' in \emph{{European
  conference for Virtual Reality and Augmented Reality}}, 2015.

\bibitem{VHI}
M.~V. Sanchez-Vives, B.~Spanlang, A.~Frisoli, M.~Bergamasco, and M.~Slater,
  ``{Virtual hand illusion induced by visuomotor correlations},'' \emph{PLoS
  One}, vol.~5, no.~4, 2010.

\bibitem{Farrer2008}
\BIBentryALTinterwordspacing
C.~Farrer, M.~Bouchereau, M.~Jeannerod, and N.~Franck, ``Effect of distorted
  visual feedback on the sense of agency,'' \emph{Behav Neurol}, vol.~19, no.
  1-2, pp. 53--57, 2008, 18413918[pmid]. [Online]. Available:
  \url{https://www.ncbi.nlm.nih.gov/pubmed/18413918}
\BIBentrySTDinterwordspacing

\bibitem{Franck2001a}
N.~Franck, C.~Farrer, N.~Georgieff, M.~Marie-Cardine, J.~Dal{\'{e}}ry,
  T.~D'Amato, and M.~Jeannerod, ``{Defective recognition of one's own actions
  in patients with schizophrenia},'' \emph{Am. J. Psychiatry}, vol. 158, pp.
  454--459, 2001.

\bibitem{Maselli2013}
A.~Maselli and M.~Slater, ``{The building blocks of the full body ownership
  illusion},'' \emph{Front. Hum. Neurosci.}, vol.~7, 2013.

\bibitem{Kokkinara2016}
E.~Kokkinara, K.~Kilteni, K.~J. Blom, and M.~Slater, ``First person perspective
  of seated participants over a walking virtual body leads to illusory agency
  over the walking,'' \emph{Scientific Reports}, vol.~6, 2016.

\bibitem{GALLAGHER200014}
S.~Ghallager, ``Philosophical conceptions of the self: implications for
  cognitive science,'' \emph{Trends Cogn. Sci.}, vol.~4, no.~1, 2000.

\bibitem{Frith2000}
C.~D. Frith, S.-J. Blakemore, and D.~M. Wolpert, ``Abnormalities in the
  awareness and control of action,'' \emph{Philosophical Transactions of the
  Royal Society of London B: Biological Sciences}, vol. 355, no. 1404, pp.
  1771--1788, 2000.

\bibitem{Wegner1999}
D.~M. Wegner and T.~Wheatley, ``{Apparent mental causation: Sources of the
  experience of will},'' \emph{Am. Psychol.}, vol.~54, no.~7, pp. 480--492,
  1999.

\bibitem{Blakemore1999}
S.~J. Blakemore, C.~D. Frith, and D.~M. Wolpert, ``{Spatio-temporal prediction
  modulates the perception of self-produced stimuli.}'' \emph{J. Cogn.
  Neurosci.}, vol.~11, no.~5, pp. 551--559, 1999.

\bibitem{Haggard2012}
P.~Haggard and V.~Chambon, ``Sense of agency,'' \emph{Current Biology},
  vol.~22, no.~10, pp. R390 -- R392, 2012.

\bibitem{Sato2005}
A.~Sato and A.~Yasuda, ``{Illusion of sense of self-agency: Discrepancy between
  the predicted and actual sensory consequences of actions modulates the sense
  of self-agency, but not the sense of self-ownership},'' \emph{Cognition},
  vol.~94, no.~3, pp. 241--255, 2005.

\bibitem{Moore2009}
J.~W. Moore, D.~M. Wegner, and P.~Haggard, ``{Modulating the sense of agency
  with external cues},'' \emph{Conscious. Cogn.}, vol.~18, no.~4, pp.
  1056--1064, 2009.

\bibitem{Wenke2010}
D.~Wenke, S.~M. Fleming, and P.~Haggard, ``{Subliminal priming of actions
  influences sense of control over effects of action},'' \emph{Cognition}, vol.
  115, no.~1, pp. 26--38, 2010.

\bibitem{Linser2007}
K.~Linser and T.~Goschke, ``{Unconscious modulation of the conscious experience
  of voluntary control},'' \emph{Cognition}, vol. 104, no.~3, pp. 459--475,
  2007.

\bibitem{Wegner2004a}
D.~M. Wegner, ``{Precis of "The Illusion of Conscious Will"},'' \emph{Behav.
  Brain Sci.}, vol.~27, pp. 649--692, 2004.

\bibitem{Moore2016}
J.~W. Moore, ``{What is the sense of agency and why does it matter?}'' p. 1272,
  2016.

\bibitem{Desantis2011}
A.~Desantis, C.~Roussel, and F.~Waszak, ``{On the influence of causal beliefs
  on the feeling of agency},'' \emph{Conscious. Cogn.}, vol.~20, no.~4, pp.
  1211--1220, 2011.

\bibitem{Haggard2002}
P.~Haggard, S.~Clark, and J.~Kalogeras, ``{Voluntary action and conscious
  awareness},'' \emph{Nat. Neurosci.}, vol. 5(4), pp. 382--385, 2002.

\bibitem{Moore2012}
J.~W. Moore and S.~S. Obhi, ``{Intentional binding and the sense of agency: A
  review},'' \emph{Conscious. Cogn.}, vol. 21(1), pp. 546--561, 2012.

\bibitem{Jeunet2018}
C.~Jeunet, L.~Albert, F.~Argelaguet, and A.~L{\'{e}}cuyer, ``{" Do you feel in
  control?": Towards Novel Approaches to Characterise, Manipulate and Measure
  the Sense of Agency in Virtual Environments},'' \emph{IEEE Trans. Vis.
  Comput. Graph.}, pp. 1--10, 2018.

\bibitem{Daprati1997}
E.~Daprati, N.~Franck, N.~Georgieff, J.~Proust, E.~Pacherie, J.~Dalery, and
  M.~Jeannerod, ``{Looking for the agent: An investigation into consciousness
  of action and self-consciousness in schizophrenic patients},''
  \emph{Cognition}, vol.~65, no.~1, pp. 71--86, 1997.

\bibitem{Kokkinara2015}
E.~Kokkinara, M.~Slater, and J.~L{\'{o}}pez-Moliner, ``{The Effects of
  Visuomotor Calibration to the Perceived Space and Body, through Embodiment in
  Immersive Virtual Reality},'' \emph{ACM Trans. Appl. Percept.}, vol.~13,
  no.~1, pp. 1--22, 2015.

\bibitem{Ma2015}
K.~Ma and B.~Hommel, ``{The role of agency for perceived ownership in the
  virtual hand illusion},'' \emph{Conscious. Cogn.}, vol.~36, pp. 277--288,
  2015.

\bibitem{Tieri2015}
G.~Tieri, E.~Tidoni, E.~F. Pavone, and S.~M. Aglioti, ``{Mere observation of
  body discontinuity affects perceived ownership and vicarious agency over a
  virtual hand},'' \emph{Exp. Brain Res.}, vol. 233, no.~4, pp. 1247--1259,
  2015.

\bibitem{Metcalfe2007}
J.~Metcalfe and M.~J. Greene, ``{Metacognition of agency},'' \emph{J. Exp.
  Psychol. Gen.}, vol. 136, no.~2, pp. 184--199, 2007.

\bibitem{Nielsen1963}
T.~I. Nielsen, ``{Volition: a new experimental approach},'' \emph{Scand. J.
  Psychol.}, vol.~4, no.~1, pp. 225--230, 1963.

\bibitem{DeVignemont2004}
F.~{De Vignemont} and P.~Fourneret, ``{The sense of agency: A philosophical and
  empirical review of the "Who" system},'' \emph{Conscious. Cogn.}, vol.~13,
  no.~1, pp. 1--19, 2004.

\bibitem{Asai2008}
T.~Asai and Y.~Tanno, ``{Highly schizotypal students have a weaker sense of
  self-agency},'' \emph{Psychiatry Clin. Neurosci.}, vol.~62, no.~1, pp.
  115--119, 2008.

\bibitem{Levenson1981}
H.~Levenson, ``{Differentiating among internality, powerful others, and
  chance},'' in \emph{Res. with Locus Control Constr.}, 1981, pp. 15--63.

\bibitem{LevensonIPC}
L.~Hanna, ``Activism and powerful others: Distinctions within the concept of
  internal-external control,'' \emph{Journal of Personality Assessment},
  vol.~38, no.~4, pp. 377--383, 1974.

\bibitem{Yokosaka2014}
T.~Yokosaka, H.~Iizuka, T.~Yonemura, D.~Kondo, H.~Ando, and T.~Maeda,
  ``Alternating images of congruent and incongruent movement creates the
  illusion of agency,'' \emph{Scientific Reports}, vol.~4, pp. 6201 EP --,
  2014.

\bibitem{Petkova2008}
V.~I. Petkova and H.~H. Ehrsson, ``{If I were you: Perceptual illusion of body
  swapping},'' \emph{PLoS One}, vol.~3, no.~12, 2008.

\bibitem{Kasahara2016}
S.~Kasahara, M.~Ando, K.~Suganuma, and J.~Rekimoto, ``{Parallel Eyes: Exploring
  Human Capability and Behaviors with Paralleled First Person View Sharing},''
  \emph{Proc. 2016 CHI Conf. Hum. Factors Comput. Syst. - CHI '16}, pp.
  1561--1572, 2016.

\bibitem{Nishida2017}
J.~Nishida and K.~Suzuki, ``{bioSync: A Paired Wearable Device for Blending
  Kinesthetic Experience},'' in \emph{Proc. 2017 CHI Conf. Hum. Factors Comput.
  Syst. - CHI '17}, 2017, pp. 3316--3327.

\bibitem{Mazzurega2011a}
M.~Mazzurega, F.~Pavani, M.~P. Paladino, and T.~W. Schubert, ``{Self-other
  bodily merging in the context of synchronous but arbitrary-related
  multisensory inputs},'' \emph{Exp. Brain Res.}, vol. 213, no. 2-3, pp.
  213--221, 2011.

\bibitem{Tsakiris2008}
M.~Tsakiris, ``{Looking for Myself: Current Multisensory Input Alters Self-Face
  Recognition},'' \emph{PLoS One}, vol. 3(12), p. e4040, 2008.

\bibitem{Sforza2010}
A.~Sforza, I.~Bufalari, P.~Haggard, and S.~M. Aglioti, ``{My face in yours:
  Visuo-tactile facial stimulation influences sense of identity},'' \emph{Soc.
  Neurosci.}, vol.~5, no.~2, pp. 148--162, 2010.

\bibitem{Tajadura-Jimenez2012a}
A.~Tajadura-Jim{\'{e}}nez, S.~Grehl, and M.~Tsakiris, ``{The other in me:
  Interpersonal multisensory stimulation changes the mental representation of
  the self},'' \emph{PLoS One}, vol.~7, no.~7, 2012.

\bibitem{Poelman2012}
R.~Poelman, O.~Akman, S.~Lukosch, and P.~Jonker, ``{As if being there: mediated
  reality for crime scene investigation},'' in \emph{Proc. ACM Conf. Comput.
  Support. Coop. Work - CSCW '12}, 2012, p. 1267.

\bibitem{Barlas2017}
Z.~Barlas, W.~E. Hockley, and S.~S. Obhi, ``{The effects of freedom of choice
  in action selection on perceived mental effort and the sense of agency},''
  \emph{Acta Psychol. (Amst).}, vol. 180, pp. 122--129, 2017.

\bibitem{Botvinick1998}
M.~Botvinick and J.~Cohen, ``Rubber hands 'feel' touch that eyes see,''
  \emph{Nature}, vol. 391, no. 6669, pp. 756--756, 1998.

\bibitem{KALCKERT2014118}
A.~Kalckert and H.~H. Ehrsson, ``The spatial distance rule in the moving and
  classical rubber hand illusions,'' \emph{Consciousness and Cognition},
  vol.~30, pp. 118 -- 132, 2014.

\bibitem{Banakou2014}
D.~Banakou and M.~Slater, ``{Body ownership causes illusory self-attribution of
  speaking and influences subsequent real speaking},'' \emph{Proc. Natl. Acad.
  Sci.}, vol. 111, no.~49, pp. 17\,678--17\,683, 2014.

\bibitem{Kalckert2012}
A.~Kalckert and H.~H. Ehrsson, ``{Moving a Rubber Hand that Feels Like Your
  Own: A Dissociation of Ownership and Agency.}'' \emph{Front. Hum. Neurosci.},
  vol.~6, p.~40, 2012.

\bibitem{Wobbrock2011}
J.~O. Wobbrock, L.~Findlater, D.~Gergle, and J.~J. Higgins, ``{The aligned rank
  transform for nonparametric factorial analyses using only anova
  procedures},'' in \emph{Proc. 2011 Annu. Conf. Hum. factors Comput. Syst. -
  CHI '11}, 2011, p. 143.

\bibitem{farrer:hal-00655447}
C.~Farrer, N.~Franck, J.~Paillard, and M.~Jeannerod, ``The role of
  proprioception in action recognition,'' \emph{Consciousness and Cognition},
  vol.~12, no.~4, pp. 609 -- 619, 2003.

\bibitem{PADRAO2016147}
\BIBentryALTinterwordspacing
G.~Padrao, M.~Gonzalez-Franco, M.~V. Sanchez-Vives, M.~Slater, and
  A.~Rodriguez-Fornells, ``Violating body movement semantics: Neural signatures
  of self-generated and external-generated errors,'' \emph{NeuroImage}, vol.
  124, pp. 147 -- 156, 2016. [Online]. Available:
  \url{http://www.sciencedirect.com/science/article/pii/S1053811915007314}
\BIBentrySTDinterwordspacing

\bibitem{Obhi2011}
S.~S. Obhi and P.~Hall, ``{Sense of agency and intentional binding in joint
  action},'' \emph{Exp. Brain Res.}, vol. 211, no. 3-4, pp. 655--662, 2011.

\bibitem{10.1371/journal.pone.0110118}
J.~A. Dewey and G.~Knoblich, ``Do implicit and explicit measures of the sense
  of agency measure the same thing?'' \emph{PLOS ONE}, vol.~9, no.~10, p.
  e110118, 2014.

\bibitem{Braun2014}
N.~Braun, J.~D. Thorne, H.~Hildebrandt, and S.~Debener, ``{Interplay of Agency
  and Ownership: The Intentional Binding and Rubber Hand Illusion Paradigm
  Combined},'' \emph{PLoS One}, vol.~9, no.~11, p. e111967, 2014.

\bibitem{Tsakiris2010}
M.~Tsakiris, M.~R. Longo, and P.~Haggard, ``{Having a body versus moving your
  body: Neural signatures of agency and body-ownership},''
  \emph{Neuropsychologia}, vol.~48, no.~9, pp. 2740--2749, 2010.

\bibitem{Tsakiris2006}
M.~Tsakiris, G.~Prabhu, and P.~Haggard, ``{Having a body versus moving your
  body: How agency structures body-ownership},'' \emph{Conscious. Cogn.},
  vol.~15, no.~2, pp. 423--432, 2006.

\bibitem{Dummer2009}
T.~Dummer, A.~Picot-Annand, T.~Neal, and C.~Moore, ``{Movement and the rubber
  hand illusion},'' \emph{Perception}, vol. 38(2), pp. 271--280, 2009.

\bibitem{Braun2018}
N.~Braun, S.~Debener, N.~Spychala, E.~Bongartz, P.~S\"{o}r\"{o}s, H.~H.~O.
  M\"{u}ller, and A.~Philipsen, ``The senses of agency and ownership: A
  review,'' \emph{Frontiers in Psychology}, vol.~9, p. 535, 2018.

\bibitem{Lecuyer2000}
A.~Lecuyer, S.~Coquillart, A.~Kheddar, P.~Richard, and P.~Coiffet,
  ``{Pseudo-haptic feedback: can isometric input devices simulate force
  feedback?}'' in \emph{Proc. IEEE Virtual Real.}, 2000, pp. 83--90.

\bibitem{Azmandian2016}
M.~Azmandian, M.~Hancock, H.~Benko, E.~Ofek, and A.~D. Wilson, ``{Haptic
  Retargeting: Dynamic Repurposing of Passive Haptics for Enhanced Virtual
  Reality Experiences},'' in \emph{Proc. 2016 CHI Conf. Hum. Factors Comput.
  Syst. - CHI '16}, 2016, pp. 1968--1979.

\bibitem{Kohli2012}
L.~Kohli, M.~C. Whitton, and F.~P. Brooks, ``{Redirected touching: The effect
  of warping space on task performance},'' \emph{IEEE Symp. 3D User Interfaces
  2012, 3DUI 2012 - Proc.}, pp. 105--112, 2012.

\bibitem{Asai2015}
T.~Asai, ``{Feedback control of one's own action: Self-other sensory
  attribution in motor control},'' \emph{Conscious. Cogn.}, vol.~38, 2015.

\bibitem{7563559}
M.~L. Chenechal, T.~Duval, V.~Gouranton, J.~Royan, and B.~Arnaldi, ``Vishnu:
  virtual immersive support for helping users an interaction paradigm for
  collaborative remote guiding in mixed reality,'' in \emph{2016 IEEE Third VR
  International Workshop on Collaborative Virtual Environments (3DCVE)}, 2016,
  pp. 9--12.

\bibitem{10.1007/978-3-642-40483-2_5}
W.~Huang, L.~Alem, and F.~Tecchia, ``{HandsIn3D: Supporting Remote Guidance
  with Immersive Virtual Environments},'' in \emph{Human-Computer Interaction
  -- INTERACT 2013}, 2013, pp. 70--77.

\bibitem{6353427}
A.~{Steed}, W.~{Steptoe}, W.~{Oyekoya}, F.~{Pece}, T.~{Weyrich}, J.~{Kautz},
  D.~{Friedman}, A.~{Peer}, M.~{Solazzi}, F.~{Tecchia}, M.~{Bergamasco}, and
  M.~{Slater}, ``Beaming: An asymmetric telepresence system,'' \emph{IEEE
  Computer Graphics and Applications}, vol.~32, no.~6, pp. 10--17, 2012.

\bibitem{Bear2011}
\BIBentryALTinterwordspacing
J.~B. Bear and A.~W. Woolley, ``The role of gender in team collaboration and
  performance,'' \emph{Interdisciplinary Science Reviews}, vol.~36, no.~2, pp.
  146--153, 2011. [Online]. Available:
  \url{https://doi.org/10.1179/030801811X13013181961473}
\BIBentrySTDinterwordspacing

\bibitem{vanBasten2009}
B.~J.~H. van Basten, S.~E.~M. Jansen, and I.~Karamouzas, ``Exploiting motion
  capture to enhance avoidance behaviour in games,'' in \emph{Motion in Games},
  A.~Egges, R.~Geraerts, and M.~Overmars, Eds.\hskip 1em plus 0.5em minus
  0.4em\relax Berlin, Heidelberg: Springer Berlin Heidelberg, 2009, pp. 29--40.

\end{thebibliography}

%

\begin{IEEEbiography}[{\includegraphics[width=1in,height=1.25in,clip,keepaspectratio]{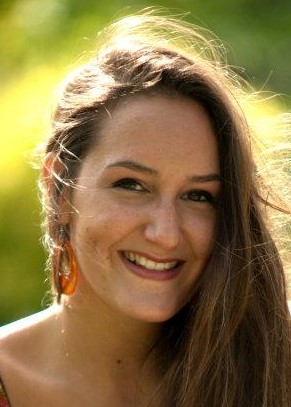}}]{Rebecca Fribourg}
graduated from the french national engineering school Universit\'e de Technologie de Compi\`egne in 2017 receiving her computer science degree. She is now a Ph.D. student at Inria Rennes, France, in Hybrid team. Her research interests include avatars and virtual embodiment. 
\end{IEEEbiography}
\vspace{-1.5cm}
\begin{IEEEbiography}[{\includegraphics[width=1in,height=1.25in,clip,keepaspectratio]{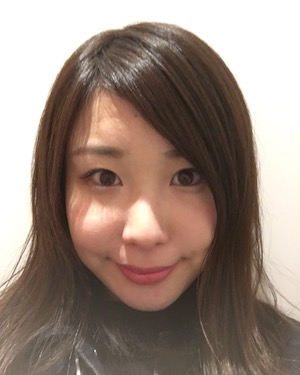}}]{Nami Ogawa}
received the M.A.Sc. degree in information studies from the University of Tokyo, Japan, in 2017. She is currently working toward the Ph.D. degree in engineering at the University of Tokyo. She contributed to this work during her research intern at Inria Rennes. Her research interests include avatar embodiment and embodied perception.
\end{IEEEbiography}
\vspace{-1.4cm}

\begin{IEEEbiography}[{\includegraphics[width=1in,height=1.25in,clip,keepaspectratio]{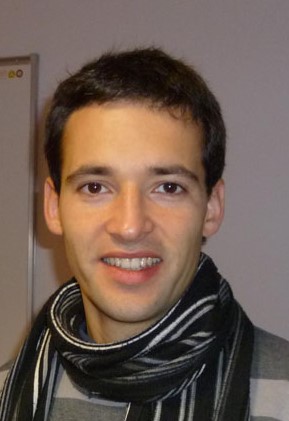}}]{Ludovic Hoyet}
is a researcher at Inria in the MimeTIC team in Rennes, France, since 2015. He received his PhD from INSA Rennes in 2010. He then worked as a Research Fellow in Trinity College Dublin under the supervision of Pr. Carol O'Sullivan. His research interests include real-time animation and perception of virtual humans.
\end{IEEEbiography}
\vspace{-1.4cm}
\begin{IEEEbiography}[{\includegraphics[width=1in,height=1.25in,clip,keepaspectratio]{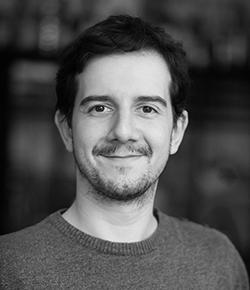}}]{Ferran Argelaguet}
is an Inria research scientist at the Hybrid team (Rennes, France) since 2016. He received his PhD degree from the Universitat Polit\`ecnica de Catalunya (UPC), in Barcelona, Spain in 2011. His main research interests include 3D user interfaces, virtual reality and human computer interaction. He is currently program co-chair of the IEEE Virtual Reality and 3D User Interfaces conference track.
\end{IEEEbiography}
\vspace{-1.4cm}
\begin{IEEEbiography}[{\includegraphics[width=1in,height=1.25in,clip,keepaspectratio]{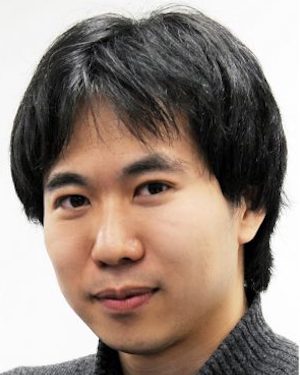}}]{Takuji Narumi}
Takuji Narumi is a lecturer at the Graduate School of Information Science and Technology, the University of Tokyo. His research interests include perceptual modification and human augmentation with virtual and augmented reality technologies. More recently, as a JST PREST researcher, he is also directing the Ghost Engineering Project which aims at utilizing the effect of virtual body on our mind and cognitive functions to encourage better communication between people.
\end{IEEEbiography}
\vspace{-1.4cm}
\begin{IEEEbiography}[{\includegraphics[width=1in,height=1.25in,clip,keepaspectratio]{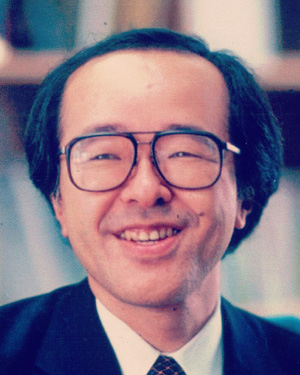}}]{Michitaka Hirose}
Michitaka Hirose is a professor of Human Interface and Virtual Reality in the Graduate school of information science and technology, the University of Tokyo. He received BE, ME, PhD in Mechanical Engineering from the University of Tokyo, in 1977, 1979, 1982, respectively. He received The 2015 VGTC Virtual Reality Career Award.
\end{IEEEbiography}
\vspace{-1.4cm}

\begin{IEEEbiography}[{\includegraphics[width=1in,height=1.25in,clip,keepaspectratio]{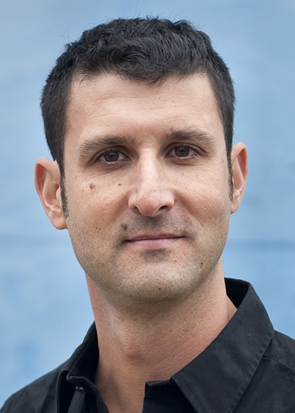}}]{Anatole L\'ecuyer}
is a senior researcher and head of Hybrid research team at Inria, the French National Institute for Research in Computer Science and Control, Rennes, France. His main research interests include virtual reality, 3D user interfaces, haptic interaction, and brain-computer interfaces. He is currently an associate editor of IEEE Transactions on Visualization and Computer Graphics, Frontiers in Virtual Environments and “Presence” journals. He was Program Chair of IEEE Virtual Reality Conference (2015-2016) and IEEE Symposium on 3D User Interfaces (2012-2013). He obtained the Inria-French Academy of Sciences Young Researcher Prize in 2013.
\end{IEEEbiography}







\end{document}